\documentclass[twocolumn]{aastex631}
\usepackage{bm}
\usepackage{amsmath}

\newcommand{\md}{\mathrm{d}}
\newcommand{\me}{\mathrm{e}}

\newcommand{\nn}{\nonumber}
\def\p{\partial}

\newcommand{\btheta}{\bm{\theta}}

\newcommand{\bJ}{\bm{J}}
\newcommand{\pattern}{\Omega_\mathrm{p}}
\newcommand{\Js}{J_\mathrm{s}}

\newcommand{\D}[2]{\frac{{\rm d} #2}{{\rm d} #1}}

\newcommand\bb[1]{\mbox{\boldmath{$#1$}}}

\newcommand\bcdot{\,\bb{\cdot}\,}

\shorttitle{Galactic bar resonances with diffusion}
\shortauthors{Hamilton et al.}
\graphicspath{{./}{figures/}}

\begin{document}

\title{Galactic bar resonances with diffusion:
\\ an analytic model with implications for bar-dark matter halo dynamical friction}

\correspondingauthor{Chris Hamilton}
\email{chamilton@ias.edu}

\author[0000-0002-5861-5687]{Chris Hamilton}
\affiliation{Institute for Advanced Study, Einstein Drive, Princeton, NJ 08540}

\author[0000-0002-2642-064X]{Elizabeth A. Tolman}
\affiliation{Institute for Advanced Study, Einstein Drive, Princeton, NJ 08540}
\affiliation{Center for Computational Astrophysics, Flatiron Institute, 162 Fifth Avenue, New York, NY 10010}

\author[0000-0002-5263-9274]{Lev Arzamasskiy}
\affiliation{Institute for Advanced Study, Einstein Drive, Princeton, NJ 08540}

\author[0000-0001-8096-7518]{Vinícius N. Duarte}
\affiliation{Princeton Plasma Physics Laboratory, Princeton University, Princeton, NJ 08543}

\begin{abstract}
The secular evolution of disk galaxies is largely driven by resonances between the orbits of `particles' (stars or dark matter) and the rotation of non-axisymmetric features (spiral arms or a bar).
Such resonances {may also explain} kinematic and photometric features observed in the Milky Way and external galaxies.
In simplified cases, these resonant interactions are well understood: for instance, the dynamics of a test particle trapped near a resonance of a steadily rotating bar is easily analyzed using the angle-action tools pioneered by Binney, Monari and others.
However, such treatments {do not} address the stochasticity and messiness 
{inherent to} real galaxies --- {effects which have, with few exceptions, been previously explored only
with complex N-body simulations}.
{In this paper}, we propose a simple kinetic equation describing the distribution function of particles near an orbital resonance with a rigidly rotating bar, allowing for diffusion of the particles' slow actions.  We solve this equation for various values of the dimensionless diffusion strength $\Delta$, {and then} apply our theory to the calculation of bar-halo dynamical friction.
For $\Delta = 0$ we recover the classic result of Tremaine \& Weinberg that friction {ultimately vanishes, owing to the phase-mixing of resonant orbits}.
However, for $\Delta > 0$ we find that
diffusion suppresses {phase-mixing}, leading to a finite torque. {Our results suggest that stochasticity --- be it physical or numerical --- tends to increase bar-halo friction, and that bars in cosmological simulations might experience significant artificial slowdown, even if the numerical two-body relaxation time is much longer than a Hubble time.}


\end{abstract}





\section{Introduction}
\label{sec:Introduction}

Galaxies are sculpted by resonances. Corotation, Lindblad and ultraharmonic resonances of rotating non-axisymmetries like bars and spirals are likely responsible for disk heating radial migration and mixing in the Milky Way \citep{Binney1988-zy, Sellwood2002-lv, Roskar2008,schonrich2009chemical,Minchev2012-ml,Sridhar2019-zo}, 
for the formation of Solar neighborhood moving groups \citep{Dehnen2000-yz,Hunt2019-qo,Kawata2021-pr},
and for 
producing the rings and dark gaps observed in external galaxies \citep{Buta1986-hq,Buta2017-we,Krishnarao2022-rh}.
The stability (or otherwise) of self-gravitating oscillation modes 
 is dictated by the
number of stars {or dark matter particles} that are able to resonate with the mode and hence {transfer angular momentum to/from it} \citep{Palmer1987-np,Weinberg1994-xq,Sellwood2014-te,Rozier2020-fc}.
Orbital resonances between stars in galactic disks, amplified by collective effects, drive rapid relaxation of the stellar distribution function (DF) \citep{Sellwood2012-ro,Fouvry2015-nk}.
The infall of heavy satellites and slowing of {galactic} bars is likely {driven by dynamical friction 
due to the resonant trapping} of dark matter particles
\citep{Lynden-Bell1972-ve,Tremaine1984-wt,Kaur2018-wp,Banik2021-ug,Chiba2021-cc}. 
Thus, the study of secular evolution of galaxies is {in large part} {the} study of resonant dynamics:
to quote from \citet{Weinberg2007-bv}, `\textit{resonances are not the exception but are required for galaxy evolution!}'

The existence of resonant structures in galaxies 
is an inevitable consequence of the quasiperiodicity of most
stellar/dark matter
orbits in the mean galactic potential. {Given this quasiperiodicity}, resonant interactions are often best described in angle-action coordinates.
\cite{Binney2012-vp,Binney2016-zh,Binney2018orbital,Binney2020-mw,Binney2020-yg} has pioneered the use of these coordinates to construct analytic equilibrium Galaxy models, and to calculate the distribution function (DF) of resonantly trapped stars.
In the latter case one assumes that the galactic potential consists of an axisymmetric mean field plus a rigidly-rotating non-axisymmetric perturbation
(see also \citealt{Monari2016-wb,Monari2017-tu,Monari2019-er}). {An} alternative way to capture the same physics is to integrate numerically the orbits of an ensemble of test particles in this prescribed potential
(e.g. \citealt{Sellwood2019-xd,Hunt2019-qo}). 
As a result of {such} efforts, resonances are now often employed in galactic dynamics as diagnostic tools: reliable dynamical features whose imprints tell us something about the underlying galaxy.  For instance, several authors have attempted to infer the size and pattern speed of the Milky Way's bar by fitting test particle models 
of its resonant imprint to Solar neighbourhood kinematic data \citep{Dehnen2000-yz,Antoja2014-vc,Trick2019-qp,Trick2021-pp}.
Even more ambitiously, \cite{Chiba2020-th,Chiba2021-cc} have suggested one might retrace the history of the Galactic bar's pattern speed by identifying a `tree ring' structure in the kinematics of {resonantly trapped} stars.  Again, the analytic and numerical tools for this analysis treated the stars as test particles under the influence of a rigidly rotating perturbation. 
A crucial element missing from these models is the diffusion of the particles in question due to stochasticity in the {gravitational} potential.  
We will refer to all models which do not include stochastic effects as \textit{collisionless}.
Stochasticity is an inevitable feature of real galaxies, arising {from} the gravitational influence of passing stars and molecular clouds, transient spiral structure, dark matter substructure, {accretion and infall from the circumgalactic environment}, or whatever \citep{Pichon2006-ak, Binney2013-cf} (to say nothing of the bar's fluctuating pattern speed and strength, overlap of resonances from other non-axisymmetric structure, and so on --- see \citealt{Minchev2012-ml,Wu2016-og,Fujii2018-tt,Daniel2019}).
Stochasticity is also inherent to simulated galaxies, e.g. because of the necessity of representing the dynamics of a very large number $N$ of stars/dark matter particles with a much smaller number of simulated particles, {which always results} in some level of numerical diffusion \citep{Weinberg2007-bv, Weinberg2007-zo, Sellwood2009-jh, ludlow2019numerical,Ludlow2021-wk, wilkinson2023impact}.  

Regardless of the source of diffusion, the implicit justification for using collisionless theory in the past {to describe resonant dynamics} has been that the libration period of particles trapped in a 
resonance, despite being much longer than the orbital period $t_\mathrm{cross}$, is still very short compared to the relaxation timescale 
$t_\mathrm{relax}$.  
The latter {quantity is defined as the timescale on which a particle's action undergoes a relative change of order unity due to stochastic effects. For instance, in the case of two-body diffusion, $t_\mathrm{relax} \sim Nt_\mathrm{cross}$.}
Yet for processes {like bar-halo friction} that depend sensitively on resonant features, the important diffusive timescale is \textit{not} the relaxation time {at all},
but rather the time to diffuse \textit{across the width of the resonance} $t_\mathrm{diff} \ll t_\mathrm{relax}$. When $t_\mathrm{diff}$ is comparable to or smaller than the libration {period}, conclusions drawn from collisionless theory may need to be revised.

The purpose of the present paper is to confront this `collisional' reality in the simplest possible model.
To do this we develop a reduced kinetic description for an ensemble of particles in the vicinity of a bar resonance, including {stochastic kicks}. In our kinetic equation, the secular part of the problem --- namely the smooth particle-bar interaction --- is treated using slow-fast angle-action variables and the pendulum approximation \citep{Chirikov1979-fj,Monari2017-tu}, while
{the stochastic effects are assumed to impact only the slow actions, and are lumped crudely into a single diffusion coefficient $D$} (which we do not attempt to calculate self-consistently).
This reduced problem then turns out to be mathematically \textit{identical} to a problem previously studied in the theory of tokamak fusion plasma devices (\citealt{BerkPPR1997,Duarte2019-mi}; see also \citealt{Pao1988-cs}).  In the tokamak context, stars are replaced with energetic ions, the galactic bar is replaced with a long-lived Alfv\'en wave, and the stochasticity stems not from molecular cloud passages or dark matter substructure but from collisions between the minority energetic ion species and the background thermal ion and electron species. 
Much like galactic dynamicists, tokamak theorists are interested in how the particle DF is heated on timescales much longer than the particles' characteristic orbital period.
What tokamaks and galaxies have in common is an underlying integrable mean-field structure that admits quasiperiodic orbits, and resonant wave-particle interactions that slowly modify those orbits. Several results can, therefore, be pulled from the tokamak literature that have not previously been employed in a stellar-dynamical context.  

Perhaps the paradigmatic example of secular particle-bar interaction is the calculation of the dynamical friction torque on a bar
embedded in a spherical dark matter halo \citep{Hernquist1992-mo,Debattista2000-dn,Athanassoula2003-xj}. 
In a classic example of collisionless theory,
 \cite{Tremaine1984-wt} analyzed this problem and found that {if one artifically freezes the bar's pattern speed}, then in the time-asymptotic limit, the DF of resonantly trapped dark matter particles phase-mixes completely, and the resulting symmetry leads to a vanishing torque on the bar (see also \citealt{Chiba2022-qt}).
As an application of our theory we revisit {this} problem here in a more general, collisional framework, and show that diffusion always injects some level of asymmetry into the DF even in {the time-asymptotic limit.
This result carries implications for the evolution of bars in both real and simulated haloes, the tension between which continues to challenge standard cosmological models \citep{roshan2021fast}.}

The rest of this paper is organised as follows.  In \S\ref{sec:Single_particle} we recap the basic angle-action formalism required to describe the secular dynamics of a single test particle near resonance in a rigidly rotating potential.  In \S\ref{sec:KT_resonance} we consider an ensemble of such particles, and write down and solve the kinetic equation that includes both this secular effect and a diffusive term.  In \S\ref{sec:Dynamical_friction} we apply what we have learned to the calculation of the dynamical friction felt by a galactic bar through coupling with its host dark matter halo.
{In \S\ref{sec:Discussion} we comment on the limitations of our theory, 
discuss implications of our results for bar evolution in both real and simulated galaxies,
and explain the relation of this work to existing literature in plasma physics.}
We summarise in \S\ref{sec:Summary}.


\section{Dynamics of a test particle near resonance}
\label{sec:Single_particle}

In this section we focus on the dynamical evolution of a test particle orbiting in a smooth galactic potential which includes a rigidly rotating bar perturbation, ignoring any stochastic effects.  We show that near a resonance the motion of the particle through phase space can be effectively reduced to that of a pendulum. 
This is a classic calculation \citep{Chirikov1979-fj} which has been employed numerous times in galactic dynamics \citep{Tremaine1984-wt,Binney2018orbital,Sridhar2019-zo}, but we repeat the key steps here in order to establish notation.
We mostly follow the notational choices of \citet{Chiba2022-qt}.

Let the gravitational potential of the galaxy be $\Phi$.
We suppose that $\Phi$ can be decomposed into a dominant, time-independent spherical background part $\Phi_0$, and a rigidly rotating bar perturbation $\delta \Phi$.
Given the spherical symmetry of $\Phi_0$ it is natural to construct a spherical coordinate system $\bm{x} \equiv (r, \vartheta, \varphi)$ which is fixed in the inertial frame with origin at the center of the galaxy.
We let the midplane of the galactic disk correspond to $\vartheta = \pi/2$, and let the 
bar be symmetric with respect to this midplane so that it rotates in the $\varphi$ direction only with constant pattern speed $\pattern$.  

We have chosen $\Phi_0$ to be spherical in order to facilitate comparison with the dynamical friction calculations of e.g. \citet{Tremaine1984-wt,Banik2021-ug,Chiba2022-qt}, where the test particles in question are dark matter particles and $\Phi_0$ represents the dark matter halo potential, but
it would be straightforward to modify our results to other contexts such as to stars orbiting in the Galactic disk \citep{Binney2018orbital}, or 
stars trapped at orbital resonances in triaxial halo potentials \citep{Yavetz2020-hj}.
In practice we will always
take our background spherical potential to be a Hernquist sphere 
\begin{equation}
    \Phi_0 = -\frac{GM}{r_\mathrm{s}+r}, 
    \label{eqn:Hernquist_potential}
\end{equation}
with total mass $M=1.5\times 10^{12} \, M_\odot$ and scale radius $r_\mathrm{s}=20$ kpc, and take the bar perturbation to be 
of the form 
\begin{eqnarray}
    &&\delta \Phi(\bm{x}, t) = \Phi_\mathrm{b}(r) \sin^2 \vartheta \cos[2(\varphi - \pattern t)],
    \label{eqn:bar_explicit}
\\
        \label{eqn:bar_details}
    && \Phi_\mathrm{b}(r) = -\frac{Av_\mathrm{c}^2}{2}\left(\frac{r}{r_\mathrm{CR}}\right)^2\left( \frac{b+1}{b+r/r_\mathrm{CR}} \right)^5,
\end{eqnarray}   
    where $r_\mathrm{CR} = v_\mathrm{c}/\Omega_\mathrm{p}$ and $A = 0.02, \, b = 0.28,\,
    v_\mathrm{c} = 238 \, \mathrm{km}\,\mathrm{s}^{-1},\, \Omega_\mathrm{p} = 35 \, \mathrm{km} \, \mathrm{s}^{-1} \, \mathrm{kpc}^{-1} = 35.8 \, \mathrm{Gyr}^{-1}$.
However, we will not need to use these explicit formulae until \S\ref{sec:numerical_results}.

Now we consider an individual test particle orbiting in the combined potential $\Phi$, and aim to describe its dynamics in the inertial frame. Its Hamiltonian is
\begin{equation}
    H = H_0 + \delta \Phi,
\end{equation}
where
\begin{equation}
     H_0 = \frac{\bm{v}^2}{2} + \Phi_0(\bm{x}),
\end{equation}
and $\bm{v}$ is the particle's velocity in the inertial frame.  The Hamiltonian $H_0$ is globally integrable, i.e. there exists a set of global angle-action coordinates $(\btheta,\bJ)$ such that $H_0$ is a function of $\bm{J}$ only.  In practice we take \citep{Binney2008-ou}:
\begin{equation}
    \btheta = (\theta_r, \theta_\psi, \theta_\varphi), \,\,\,\,\,\,\,\,\,\,\, \bJ = (J_r, L, L_z).
\end{equation}
The angles $\bm{\theta}$
quantify the phase 
of oscillations in the radial direction, the phase of oscillations in the azimuthal direction within the particle's orbital plane, and the longitude of ascending node of the particle's orbit (which is fixed for spherical $\Phi_0$) respectively.  The corresponding actions $\bJ$ quantify the amplitude of radial excursions, the total angular momentum, and the $z$-component of angular momentum of the particle's orbit.
In the limit of a vanishingly weak bar perturbation $\delta \Phi\to 0$, the Hamiltonian $H = H_0$ only depends on $\bJ$; in this case the actions $\bJ$ are perfectly conserved while the angles $\btheta$ evolve linearly as $\bm{\theta}(t) = \bm{\theta}(0) + \bm{\Omega}({\bJ})t$ where
\begin{equation}
    \boldsymbol{\Omega}(\bJ) \equiv \frac{\partial H_0}{\partial \bJ} = (\Omega_r, \, \Omega_\psi, \, \Omega_\varphi),
\end{equation}
is the vector of orbital frequencies of the unperturbed problem.  Since {$\theta_\varphi$} is fixed for a particle orbiting a spherical potential we always have $\Omega_\varphi = 0$. 

Canonical Hamiltonian perturbation theory allows us to describe the modification of orbits by the finite bar perturbation $\delta \Phi$ \citep{Binney2008-ou}.  
 However, canonical theory breaks down near orbital resonances, namely locations $\bJ_\mathrm{res}$ in action space such that
 \begin{equation}
     \bm{N}\bcdot \bm{\Omega}(\bJ_\mathrm{res}) = N_\varphi \pattern,
     \label{eqn:resonance_condition}
 \end{equation}
for some vector of integers $\boldsymbol{N} = (N_r, N_\psi, N_\varphi)$.
To describe orbits in the vicinity of such resonances it is best to make a canonical transformation to a new set of coordinates, which are again angle-action coordinates of the unperturbed problem (\citealt{Lichtenberg2013-pw,Binney2020-mw}).
Precisely, we map
$(\btheta, \bJ) \to (\btheta',\bJ')$, where $\btheta' = (\btheta_\mathrm{f}, \theta_{\mathrm{s}})$ consists of the `fast' and `slow' angles\footnote{{Formally, this mapping follows from the 
time-dependent generating function $s(\btheta,\bJ',t) = \theta_r J_{\mathrm{f}1} + \theta_\psi J_{\mathrm{f}2}
+ (\bm{N}\cdot\btheta - N_\varphi \pattern t)J_\mathrm{s}$}} 
\begin{eqnarray}
\bm{\theta}_{\mathrm{f}} &=& (\theta_{\mathrm{f}1}, \theta_{\mathrm{f}2}) \equiv \left( \theta_r, \theta_\psi \right),
\label{eqn:fast_angles}
\\
\theta_s &\equiv& \bm{N}\bcdot \boldsymbol{\theta} - N_\varphi \pattern t,
\label{eqn:slow_angle}
\end{eqnarray}
and $\bJ' = (\bJ_{\mathrm{f}}, J_\mathrm{s})$
consists of the corresponding fast and slow actions
\begin{align}
 \bm{J}_{\mathrm{f}} &= (J_{\mathrm{f}1}, J_{\mathrm{f}2}) \equiv \left( J_r- \frac{N_r}{N_\varphi} L_z ,\,\,\,\, \, L- \frac{N_\psi}{N_\varphi}L_z \right),
\label{eqn:fast_actions}
\\
J_\mathrm{s} &\equiv L_z/N_\varphi.
\label{eqn:slow_action}
\end{align}
Having made this transformation we may rewrite $H$ in terms of the new coordinates. It takes the form
\begin{eqnarray}
    H(\btheta',\bJ') &=& H_0(\bJ')  - N_\varphi \pattern J_\mathrm{s} 
    \nn \\
    && + \sum_{\bm{k}\neq \bm{0}} \Psi_{\bm{k}}(\bJ') \exp(i\bm{k}\bcdot \bm{\theta}'),
    \label{eqn:Hamiltonian_slowfastangles}
\end{eqnarray}
where $\bm{k} = (k_1, k_2, k_\mathrm{s})$ is a vector of integers and 
we have expanded $\delta \Phi$ as a Fourier series in the new angles $\btheta'$, i.e.
written $\delta \Phi(\bm{x}, t) = \sum_{\bm k} \Psi_{\bm k}(\bJ') \exp(i\bm{k}\bcdot \btheta')$.  The coefficients $\Psi_{\bm k}$ are easily computed for the simple bar model \eqref{eqn:bar_explicit} --- see \cite{Tremaine1984-wt} and Appendix B of \cite{Chiba2022-qt}.
The special thing about the form (\ref{eqn:Hamiltonian_slowfastangles}) of the Hamiltonian is that it has no explicit time dependence (or rather, the time-dependence has been absorbed into the definition of the angle $\theta_\mathrm{s}$; see equation \eqref{eqn:slow_angle}).

The fast angles $\btheta_\mathrm{f}$ evolve on the orbital timescale whereas $\theta_\mathrm{s}$ evolves on the much longer timescale $\sim (\bm{N}\bcdot \bm{\Omega} - N_\varphi \pattern)^{-1}$.  Thus we may average $H$ over the unimportant fast motion; the result is
\begin{eqnarray}
    \mathcal{H}  &\equiv& \frac{1}{(2\pi)^2}\int \md \btheta_\mathrm{f} \, H(\btheta', \bJ') \nn 
    \\
    &=&  H_0(J_\mathrm{s}) - N_\varphi \pattern \Js
    +
    \sum_{k\neq 0} \Psi_{k}(J_\mathrm{s}) \exp(ik {\theta}_\mathrm{s}),
    \label{eqn:averaged_H}
 \end{eqnarray}
 where we used the shorthand $\Psi_{(0,0,k)} = \Psi_k$, and
 the dependence on fast actions $\bJ_\mathrm{f}$ is now implicit.
Hamilton's equations tell us that $\bJ'$ evolves according to $\md \bJ'/\md t = - \partial \mathcal{H}/\partial \btheta' = (-\partial \mathcal{H}/\partial \theta_\mathrm{s}, 0, 0)$.
Thus the fast actions $\bJ_\mathrm{f}$ are constants under the time-averaged perturbation, and so at a fixed $\bJ_\mathrm{f}$ we find that the dynamics reduces to motion in the `slow plane'
  $(\theta_\mathrm{s}, J_\mathrm{s})$.
We can simplify $\mathcal{H}$ further by exploiting the fact that we only care about motion in the vicinity of the resonance (indeed, we already made this restriction implicitly by assuming that $\theta_\mathrm{s}$ evolves much more slowly than $\btheta_\mathrm{f}$).
To do this, let the resonant action $\bJ_\mathrm{res}$
from equation (\ref{eqn:resonance_condition}) transform to $\bJ'_\mathrm{res} = (\bJ_\mathrm{f}, J_\mathrm{s, res})$, and define
\begin{equation}
    I \equiv J_\mathrm{s} - J_{\mathrm{s, res}} , \,\,\,\,\,\,\,\,\,\,\, 
    \phi_k = \theta_\mathrm{s} + \frac{\arg \Psi_k}{k},
    \label{eqn:I_phi_definitions}
\end{equation}
and let $\phi_k \in (-\pi, \pi)$.
Then Taylor expanding the right hand side of \eqref{eqn:averaged_H} for small $I$ and discarding constants and unimportant higher order terms, we find\footnote{For a discussion of the accuracy of the approximations made in deriving equation \eqref{eqn:averaged_H_expanded}, and the leading-order corrections to it, see \cite{Kaasalainen1994-ks}, \cite{Binney2018orbital, Binney2020-yg}.}
 \begin{eqnarray}
     \mathcal{H} &\approx& \frac{1}{2} H_0''(J_\mathrm{s, res})I^2 +
    2 \sum_{k > 0}  \vert \Psi_k(J_\mathrm{s,res}) \vert \cos (k \phi_k).
    \label{eqn:averaged_H_expanded}
 \end{eqnarray}

Normally it is the case that the sum over $k$ in equation \eqref{eqn:averaged_H_expanded} is dominated by a single, small value of $k$.
Here we will focus on resonances $\bm{N}$ with $N_\varphi = 2$, since these give the dominant contributions to the dynamical friction on galactic bars, although the formalism we develop can be applied to other resonances with minimal adjustments.  For $N_\varphi=2$ the only contribution to equation \eqref{eqn:averaged_H_expanded} is from $k=1$ (\citealt{Chiba2022-qt}, Appendix B).
The resulting simplified $\mathcal{H}$ therefore takes the pendulum form \citep{Chirikov1979-fj,Lichtenberg2013-pw}:
\begin{equation}
    \mathcal{H} = \frac{1}{2}GI^2 - F\cos \phi,
    \label{eqn:resonant_Hamiltonian}
\end{equation}
where $\phi \equiv \phi_1$ and
\begin{equation}
    G \equiv \frac{\partial^2 H_0}{\partial J_\mathrm{s}^2}\Bigg\vert_{J_{\mathrm{s,res}}}, \,\,\,\, \,\,\,\,\,\, \,\,\,\,
    F = -2\vert \Psi_1(J_{\mathrm{s,res}})\vert .
    \label{eqn:F_and_G}
\end{equation}
In the cases of interest to us $G<0$, and of course $F<0$ always.
An explicit expression for $-F$ is given in equation (B9) of \citet{Chiba2022-qt}. 

The variables $(\phi, I)$ are canonical variables for the pendulum Hamiltonian \eqref{eqn:resonant_Hamiltonian}.
Their equations of motion are
\begin{equation}
    \frac{\md \phi}{\md t} = \frac{\partial \mathcal{H}}{\partial I} = GI,\,\,\,\,\,\,\,\,
    \,\,\,\,\,
    \frac{\md I}{\md t} = -\frac{\partial \mathcal{H}}{\partial \phi} = -F\sin \phi,
\end{equation}
which are the equations of a simple pendulum.  The pendulum moves at constant `energy' $\mathcal{H}$,  either on an untrapped `circulating' orbit with $\mathcal{H} > F$ (so that $\phi$ periodically takes all values in $(-\pi, \pi)$), or on a trapped `librating'  orbit with $\mathcal{H} < F$ (so that the pendulum oscillates around $(\phi, I) = (0,0)$).  The separatrix between the two orbit families has the equation $\mathcal{H} = F$.
The {period of} infinitesimally small oscillations around $(\phi, I) = (0,0)$ --- the so-called \textit{libration time} --- is given by
\begin{equation}
    t_\mathrm{lib} \equiv \frac{2\pi}{\sqrt{FG}}.
    \label{eqn:t_libration}
\end{equation}
The maximum width in $I$ of the librating `island' is at $\phi = 0$, where it spans $I\in (-I_\mathrm{h}, I_\mathrm{h})$ and $I_\mathrm{h}$ is the \textit{island half-width}:
\begin{equation}
    I_\mathrm{h} \equiv 2\sqrt{\frac{F}{G}}.
    \label{eqn:half_width}
\end{equation}
Equations (\ref{eqn:t_libration}) and (\ref{eqn:half_width}) are the fundamental scales involved in the dynamics of test particles governed
by the pendulum Hamiltonian (\ref{eqn:resonant_Hamiltonian}).



\section{Kinetic theory near a resonance}
\label{sec:KT_resonance}

In \S\ref{sec:Single_particle} we learned how to describe the dynamical evolution of a single resonant test particle in the smooth, rigidly rotating potential $\Phi_0+\delta \Phi$.  In this section we will add a stochastic forcing to the dynamics.  The motion of a single particle is no longer deterministic in this case, so we
must turn to a statistical description of an ensemble of such test particles.  We do this using a kinetic equation.

We wish to understand how an ensemble of particles 
behaves near a particular resonance $\bm{N}$ associated with the slow action $J_{s,\mathrm{res}}$.
Let us therefore consider particles that all share the same fast actions
$\bJ_\mathrm{f}$, but may differ in their values of fast angles $\btheta_\mathrm{f}$, slow angle $\theta_\mathrm{s}$ and slow action $\Js$.
Averaging the ensemble over the fast angles, we can describe the resulting density of particles in the $(\phi, I)$ plane (equation (\ref{eqn:I_phi_definitions})) using a smooth distribution function (DF)
which we call $f(\phi, I, t)$. The number of particles in the phase space area element $\md \phi \, \md I$ surrounding $(\phi, I)$ at time $t$ is then proportional to
$\md \phi\,  \md I f(\phi, I, t)$.
The kinetic equation governing $f$ is
\begin{equation}
    \frac{\partial f}{\partial t} + \{ f, \mathcal{H}\} = C[f],
    \label{eqn:kinetic_equation_general}
\end{equation}
where $\{\cdot,\cdot\}$ is a Poisson bracket encoding the smooth advection in the $(\phi, I)$ plane:
\begin{equation}
    \{ f,\mathcal{H}\} = \frac{\partial f}{\partial \phi} \frac{\partial \mathcal{H}}{\partial I} - \frac{\partial f}{\partial I} \frac{\partial \mathcal{H}}{\partial \phi},
    \label{eqn:Poisson_bracket}
\end{equation}
and $C[f]$ is a `collision operator' encoding the effect of stochastic fluctuations in the potential. For $\mathcal{H}$ we use the pendulum approximation (\ref{eqn:resonant_Hamiltonian}), while for $C[f]$ we take a very simple diffusive form with constant diffusion coefficient $D$:
\begin{equation}
    C[f] = D \frac{\partial^2 f}{\partial I^2}.
    \label{eqn:collision_operator}
\end{equation}
The diffusion coefficient $D$ can be {calculated from} a given theoretical model of e.g. scattering by passing stars and molecular clouds \citep{Binney1988-zy,Jenkins1990-qs,S_De_Simone2004-gr}, dark matter substructure \citep{Penarrubia2019-su, El-Zant2020-oc, Bar-Or2019-ge}, 
spurious numerical heating in simulations \citep{Weinberg2007-bv,Ludlow2021-wk}, or can potentially be calibrated from data \citep{Ting2019-la,Frankel2020-vy}. {See \S\ref{sec:Delta} for some quantitative estimates}.
We emphasize that we make no attempt at self-consistency here, as we simply impose a diffusion coefficient by hand, rather than calculating it from $f$.  Similarly, we do not account at any stage for the self-gravity of the perturbed {dark matter DF; in other words, the particles interact only with the bar, not with each other \citep{Weinberg1985-en,Dootson2022-cu}}.

Putting equations \eqref{eqn:resonant_Hamiltonian} and \eqref{eqn:kinetic_equation_general}--\eqref{eqn:collision_operator} together, we have {the kinetic equation}:
\begin{equation}
    \frac{\partial f}{\partial t} + GI\frac{\partial f}{\partial \phi} - F\sin \phi\frac{\partial f}{\partial I}= D\frac{\partial^2 f}{\partial I^2}.
    \label{eqn:kinetic_equation}
\end{equation}
In writing down {equation} \eqref{eqn:kinetic_equation} we have made a few key assumptions:
\begin{itemize}
    \item We have ignored any diffusion in $\phi$.
    \item We have assumed that the diffusion coefficient $D$ is a constant, independent of $\phi$ or $I$.
    \item We have assumed that it is legitimate to consider an ensemble at fixed fast action $\bJ_\mathrm{f}$, which implicitly ignores any diffusion in $\bJ_\mathrm{f}$. 
\end{itemize}
We can justify the first two assumptions heuristically on the grounds that the resonant island is \textit{narrow}, i.e. the extent of the island in $I$ is usually small compared to the whole of slow action space, whereas it spans the whole of slow angle $(\phi)$ space.\footnote{This `narrow resonance approximation' is standard in tokamak plasma theory \citep{Su1968-ga,Callen, Duarte2019-mi,Tolman2021-ly}. 
} As a result, the features in the DF created by the resonance are going to be much sharper in $I$ than in $\phi$, so a term like $\partial^2 f/\partial \phi^2$ in the collision operator is expected to be sub-dominant. 
Also, the narrowness of the island means we are only interested in the dynamics over a small portion of action space, so $D$ ought not to vary too much across the domain of interest and can be estimated locally.
The third assumption, namely that of no diffusion in $\bJ_\mathrm{f}$, is difficult to justify in the general case ---
there is a priori no reason to assume that stochastic kicks in $\bJ_\mathrm{f}$ will be small compared to those in $J_\mathrm{s}$, and such kicks will couple different $\bJ_\mathrm{f}$ in a proper kinetic equation via terms like $\sim \partial^2 f/\partial J_{\mathrm{f}1}^2$.  Nevertheless, we will stick with this assumption because it 
allows a great simplification of the kinetic theory, and the basic results and intuition we gain from solving the simplified theory will be invaluable when attacking more realistic problems. We discuss this assumption further in \S\ref{sec:Discussion}.



Before attempting to solve the kinetic equation \eqref{eqn:kinetic_equation}, one more simplification is in order.
Naively, equation \eqref{eqn:kinetic_equation} seems to depend on three parameters: $G, F$ and $D$.  However, we can reduce this to one effective parameter by introducing certain dimensionless variables.
First we note that the typical timescale for a particle to diffuse all the way across the resonance (a distance $\sim 2I_\mathrm{h}$ in action space) is the \textit{diffusion time}, 
\begin{equation}
    t_\mathrm{diff} \equiv \frac{(2I_\mathrm{h})^2}{2D}= \frac{8F}{GD}.
\label{eqn:diffusion_time}
\end{equation}
Now let us define the dimensionless variables
\begin{eqnarray}
\tau &\equiv& \sqrt{GF}t = \frac{2\pi \,t}{t_\mathrm{lib}},
\label{eqn:dimensionless_time}
   \\
   j &\equiv& \sqrt{\frac{G}{F}}I = \frac{2I}{I_\mathrm{h}},
   \label{eqn:dimensionless_action}
  \\
  \Delta &\equiv& \sqrt{\frac{G}{F^3}}D = \frac{4\, t_\mathrm{lib}}{\pi\, t_\mathrm{diff}},
   \label{eqn:dimensionless_diffusion}
\end{eqnarray}
where the libration time $t_\mathrm{lib}$ is given in equation \eqref{eqn:t_libration} and the island half-width $I_\mathrm{h}$ is defined in equation \eqref{eqn:half_width}.
Clearly, $\tau$ is a dimensionless measure of time normalized by the libration time, $j$ is a dimensionless measure of the slow action variable relative to the resonant island width, and $\Delta$ is the \textit{diffusion strength}, i.e. the ratio of libration timescale to diffusion timescale.
Treating $f$ as a function of these variables, i.e. writing\footnote{Note that $f$ still depends parametrically on $\bm{N}$ and $\bJ_\mathrm{f}$, though we are suppressing the explicit dependence here to keep the notation clean.} $f(\phi, j,\tau)$, equation (\ref{eqn:kinetic_equation}) becomes:
\begin{equation}
    \frac{\partial f}{\partial \tau} + j\frac{\partial f}{\partial \phi} - \sin \phi\frac{\partial f}{\partial j}= \Delta \frac{\partial^2 f}{\partial j^2}.
    \label{eqn:kinetic_equation_dimensionless}
\end{equation}
We see that in these variables the kinetic equation depends on the single dimensionless parameter $\Delta$, which we discuss next.

\subsection{The diffusion strength $\Delta$}
\label{sec:Delta}

The dimensionless diffusion strength $\Delta \approx t_\mathrm{lib}/t_\mathrm{diff}$ (equation \eqref{eqn:dimensionless_diffusion}) plays a central role in this work, since it measures the importance of diffusion relative to that of the bar perturbation near resonance.
The regime $\Delta \gg 1$ corresponds to very strong diffusion,
whereas 
$\Delta \ll 1$ corresponds to very weak diffusion; the `collisionless limit' favored by many classic calculations corresponds to $\Delta =0$ (see \S\ref{sec:Introduction}). 
A key aim of this paper is to understand how the behavior of a DF near a resonance
depends on $\Delta$, and how this might affect galactic phenomena like bar-halo friction.
As we will see, even for $\Delta$ values as small as $\sim 0.1$ the behavior predicted by the kinetic equation \eqref{eqn:kinetic_equation_dimensionless} can be significantly different from the collisionless limit.

We can estimate $\Delta$ heuristically by noting that
$t_\mathrm{diff}$ is related to the relaxation time $t_\mathrm{relax}$, which is the time required for {any stochastic process} to change a particle's action by order of itself \citep{Binney2008-ou,Fouvry2018-gi}.  
If we assume the diffusion coefficient $D$ is roughly constant over the whole of action space (which is a gross oversimplification but will suffice for an order-of-magnitude estimate), then from equation \eqref{eqn:diffusion_time} we have:
\begin{eqnarray}
\frac{t_\mathrm{diff}}{t_\mathrm{relax}} \sim \left(\frac{I_\mathrm{h}}{J_\mathrm{s, res}}\right)^2
\sim \bigg\vert \frac{\Psi_1}{H_0} \bigg\vert \sim A \ll 1.
\label{eqn:ratio_diff_relax}
\end{eqnarray}  
Here $A$ is the dimensionless strength of the bar (equation \eqref{eqn:bar_details}); 
for the Milky Way, $A \sim 0.02$ \citep{Chiba2020-th}.
Putting equations (\ref{eqn:dimensionless_diffusion}) and (\ref{eqn:ratio_diff_relax}) together, we estimate 
\begin{eqnarray}
\Delta &\sim& \frac{t_\mathrm{lib}}{A t_\mathrm{relax}} 
\label{eqn:Delta_approximate}
\\
&\sim& 10 \times
\left( \frac{A}{0.02} \right)^{-1}
\left( \frac{t_\mathrm{lib}}{2\, \mathrm{Gyr}} \right)
\left( \frac{t_\mathrm{relax}}{10\, \mathrm{Gyr}} \right)^{-1}. \label{eqn:Delta_numerical}
\end{eqnarray}  
Typically, important Galactic bar resonances like the corotation resonance have libration times of $t_\mathrm{lib} \sim 2$ Gyr (see \S\ref{sec:Dynamical_friction} for examples). 
{The relaxation time $t_\mathrm{relax}$ depends on
the context
in which the kinetic equation is being applied, since diffusion can be driven by a variety of different causes, including 
transient spiral arms, 
finite-$N$ granularity noise, etc.}
{But regardless of the context, we see immediately from the estimate 
\eqref{eqn:Delta_numerical} that $\Delta$ can be of order unity \textit{even for systems with relaxation times much longer than a Hubble time}.} 

{To be more quantitative, let us consider 
a virialized, isolated dark matter halo of enclosed mass $M(r)$ made up of particles of mass $m$ and velocity dispersion $\sigma(r)$. Then the two-body relaxation time for the dark matter particles is \citep{Binney2008-ou}:
\begin{equation}
    t_\mathrm{relax} \sim \frac{\sigma^3 r^3}{G^2 m M} \sim N t_\mathrm{cross},
    \label{eqn:2body}
\end{equation}
where $N = M/m$ is the number of particles inside radius $r$,
and we used $\sigma^2 \sim GM/r$ and introduced the typical crossing time  $t_\mathrm{cross} \sim \sigma/R$.
Moreover, a reasonable estimate for the libration time is $t_\mathrm{lib} \sim A^{-1/2}t_\mathrm{cross}$.
Plugging these results in to \eqref{eqn:Delta_approximate} gives
\begin{equation}
    \Delta \sim \frac{1}{A^{3/2} N}
\approx 0.1 \times
\left( \frac{A}{0.01} \right)^{-1}
\left( \frac{N}{10^4} \right)^{-1},
\label{eqn:Delta_N}
\end{equation}
where the quantities on the right hand side should be evaluated 
at the location of the important resonances (usually $r\lesssim 10$ kpc). 
For standard cold dark matter models, $N$ is enormous, which results in a 
negligible $\Delta \lll 1$, rendering collisionless theory valid. 
However, in numerical simulations of cold dark matter haloes, $N$ is the number of particles used to represent the halo inside the resonant radius, which is often comparable to $10^4$ or $10^5$ within $10$ kpc \citep{ludlow2019numerical,Ludlow2021-wk}.
It follows that numerical noise alone can produce an effective $\Delta \sim 0.1$.
Similarly, $\Delta$ may be non-negligible in alternative dark matter models; for instance in 
fuzzy dark matter theory, particles with true masses $m \sim 10^{-22}$ eV behave collectively as quasiparticles of size $\sim 1$ kpc and effective mass $m_\mathrm{eff} \gtrsim 10^7 M_\odot$ (see \citealt{Hui2017-rw}, equation (38)), so 
a halo of enclosed mass $M=10^{11}M_\odot$ would behave as if it were constituted of $N\sim 10^4$ particles,
again giving $\Delta \sim 0.1$.}

As another example, we might let our DF $f$ describe not dark matter particles, but \textit{stars} trapped near bar resonances in the Milky Way disk
\citep{Monari2017-tu,Trick2019-qp, Dootson2022-cu}.  In this case one can set  $t_\mathrm{relax}$ equal to the typical diffusion timescale for the orbital actions of disk stars, as measured from kinematic and chemical data from GAIA and APOGEE
(i.e. the `age-velocity dispersion relation', see \citealt{Mackereth2019-cq,Frankel2020-vy}). These measurements suggest that $t_\mathrm{relax}$ is on the order of the age of the Galaxy;
it follows from equation \eqref{eqn:Delta_approximate} that typical $\Delta$ values are $\gtrsim 1$. 
Of course, such estimates ignore the fact that the bar resonances themselves likely have some influence on the diffusion of stars, particularly when they overlap with other bar resonances and/or resonances associated with spiral waves \citep{Minchev2010-la}, meaning it may be unrealistic to decouple secular and stochastic effects like we have proposed here.
We assume throughout this paper that resonances are isolated,
so we can make no quantitative statement here about the effect of resonance overlap.
We leave this as an avenue for careful future study.



\subsection{{Solving} the kinetic equation}
\label{sec:solution}

Unfortunately, the kinetic equation (\ref{eqn:kinetic_equation_dimensionless}) cannot be solved analytically in general. 
In this section we present time-dependent numerical solutions to (\ref{eqn:kinetic_equation_dimensionless}) for a simple choice of initial DF and for various values of $\Delta$. In particular we find that the solution always tends towards some steady-state form.
There are two interesting classes into which the steady-state solutions fall --- for $\Delta \ll 1$ the steady DF is dominated by the resonant island structure, while for $\Delta \gg 1$ it is dominated by diffusion and the resonance leaves little trace.  
In fact, these limiting cases have already been investigated analytically in the context of
wave-particle interactions in plasma \citep{Pao1988-cs,Berk1998-hv,Duarte2019-mi}.
Inspired by these works, in \S\ref{sec:weak_scattering} and
\S\ref{sec:strong_scattering} we derive some analytical solutions to (\ref{eqn:kinetic_equation_dimensionless})
in the $\Delta \ll 1$ and $\Delta \gg 1$ limits respectively. We will refer to those solutions throughout this section in order to illuminate our numerical results.


Let the initial global distribution of particles be a function of unperturbed actions $\bJ$ only, so by Jeans' theorem the initial `background' DF $f_0(\bJ)$ --- normalized so that $\int \md \btheta \md \bJ \, f_0 = 1$ --- is a steady-state solution of the unperturbed 
collisionless Boltzmann equation governed by $H_0(\bJ)$ \citep{Binney2008-ou}. 
Since the canonical transformation $(\btheta, \bJ) \to (\btheta',\bJ')$ does not mix up angles and actions (equations \eqref{eqn:slow_angle}--\eqref{eqn:fast_actions}), it follows that when written in the new coordinates $f_0$ will just be a function of $\bJ'$.
In the language of the dimensionless variables we used to write down equation \eqref{eqn:kinetic_equation_dimensionless}, at fixed $\bm{J}_\mathrm{f}$ the initial DF in the $(\phi, j)$ plane will depend only on $j$, so $f_0 = f_0(j)$.

For simplicity, we will only focus on DFs which are initially \textit{linear} in $j$.
This is well-motivated provided the resonance is narrow (\S\ref{sec:KT_resonance}) since any initial DF which depends only on $j$ can be Taylor-expanded around the resonance at $j=0$ to give a locally linear form.
Thus we take as our initial condition
\begin{equation}
    f(\phi, j, \tau=0) = f_0(j=0) + \alpha j \equiv f_\mathrm{init}(j),
    \label{eqn:linear_DF}
\end{equation}
where
\begin{equation}
    \alpha \equiv \frac{\partial f_0}{\partial j}\bigg\vert_{j=0} 
    .
\end{equation}
Of course the linear approximation (\ref{eqn:linear_DF}) should be corrected in detailed modelling \citep{Pao1988-cs} but it is sufficient to elucidate the main physics of equation (\ref{eqn:kinetic_equation_dimensionless}) \citep{White2019-nl}.

\begin{figure}
\centering
\includegraphics[width=0.49\textwidth]{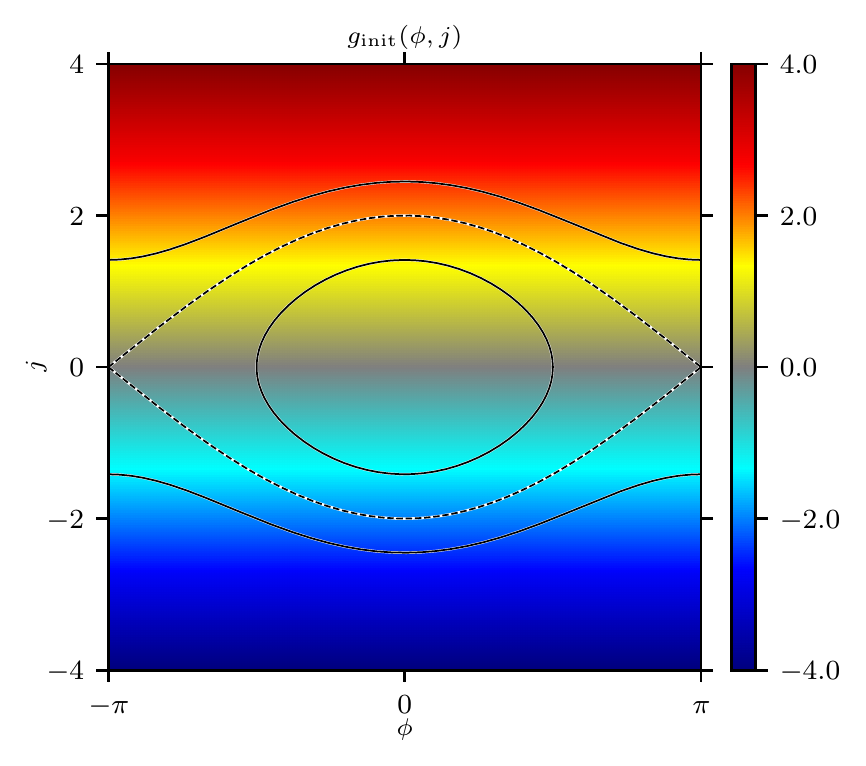}
\caption{Colors show contours of the initial condition $g_{\mathrm{init}}(\phi, j) \equiv g(\phi, j, 0)= j$, where $g$ is related to the DF $f$ through equation \eqref{eqn:dimensionless_DF}. We overlay with solid black lines contours of the pendulum Hamiltonian $\mathcal{H}$. The separatrix between librating and circulating regions is shown with a black dashed line.}
\label{fig:dimensionless_DF}
\end{figure}
At $t=0$ we switch on the bar potential, and the resulting evolution is described by equation (\ref{eqn:kinetic_equation_dimensionless}).
Of course this is not realistic --- true galactic bars do not just appear out of nowhere but instead grow gradually in strength in the first few Gyr of the galaxy's lifetime.
However, the qualitative results one finds by growing the bar slowly are not too different from imposing it instantaneously at $t=0$ \citep{Chiba2022-qt}\footnote{In fact \citet{Chiba2022-qt} show that in the collisionless regime the DF exhibits sharper features {near the separatrix} when the bar is grown slowly than when it is imposed instantaneously.  Since diffusion feeds off sharp features in the DF (equation (\ref{eqn:collision_operator})), diffusive effects may be even more important in this case.}. For simplicity we will consider only an instantaneously emerging bar here, {but time-dependent bar strengths (and pattern speeds) will be considered in future work}.

\subsubsection{{Numerical method}}
\label{sec:numerical_method}

{We solve equation \eqref{eqn:kinetic_equation_dimensionless} numerically on a regular grid in $(\phi, j)$ space, 
with $200$ cells in $\phi$ (from $\phi_{\rm min} = -\pi$ to $\phi_{\rm max} = \pi$) and $4000$ cells in $j$ (from $j_{\rm min} = -20\pi$ to $j_{\rm max} = 20\pi$). For the integration we use an explicit forward-Euler scheme, which is first-order accurate in $\tau$, and second-order accurate in $\phi$ and $j$.
We also set $f(\phi,j_{\rm  min,max},\tau) = f_\mathrm{init}(\phi,j_{\rm  min,max})$ in order to satisfy
the boundary condition that $f \to f_\mathrm{init}$ for very large $\vert j \vert$ (see \S\ref{sec:angle_averaged}).
Because we are solving on a grid we inevitably have some numerical diffusion, which makes it hard to resolve the smallest-scale structures which arise at late times for very small $\Delta$; to avoid this issue,
we report numerical results only for $\Delta \geq 0.001$. The case $\Delta = 0$ can be solved analytically \citep{Chiba2022-qt}.
}

\subsection{{Results}}

We can most easily examine the nature of the resulting solutions by defining a dimensionless auxiliary DF:
\begin{equation}
    g(\phi, j, \tau) \equiv \frac{f(\phi, j, \tau) - f_0(0)}{\alpha},
    \label{eqn:dimensionless_DF}
\end{equation}
which measures the modification of $f$ compared to its background value on resonance $f_0(0)$; it is obvious from \eqref{eqn:linear_DF} that $g(\phi, j, \tau=0) \equiv g_\mathrm{init}(j) =  j$.  Figure \ref{fig:dimensionless_DF} shows colored contours of this initial distribution, overlaid by black contours of the Hamiltonian $\mathcal{H}$ (equation \eqref{eqn:resonant_Hamiltonian}).  The separatrix between librating and circulating orbit families is shown with a 
dashed black line.
\begin{figure*}
\centering
\includegraphics[width=\textwidth]{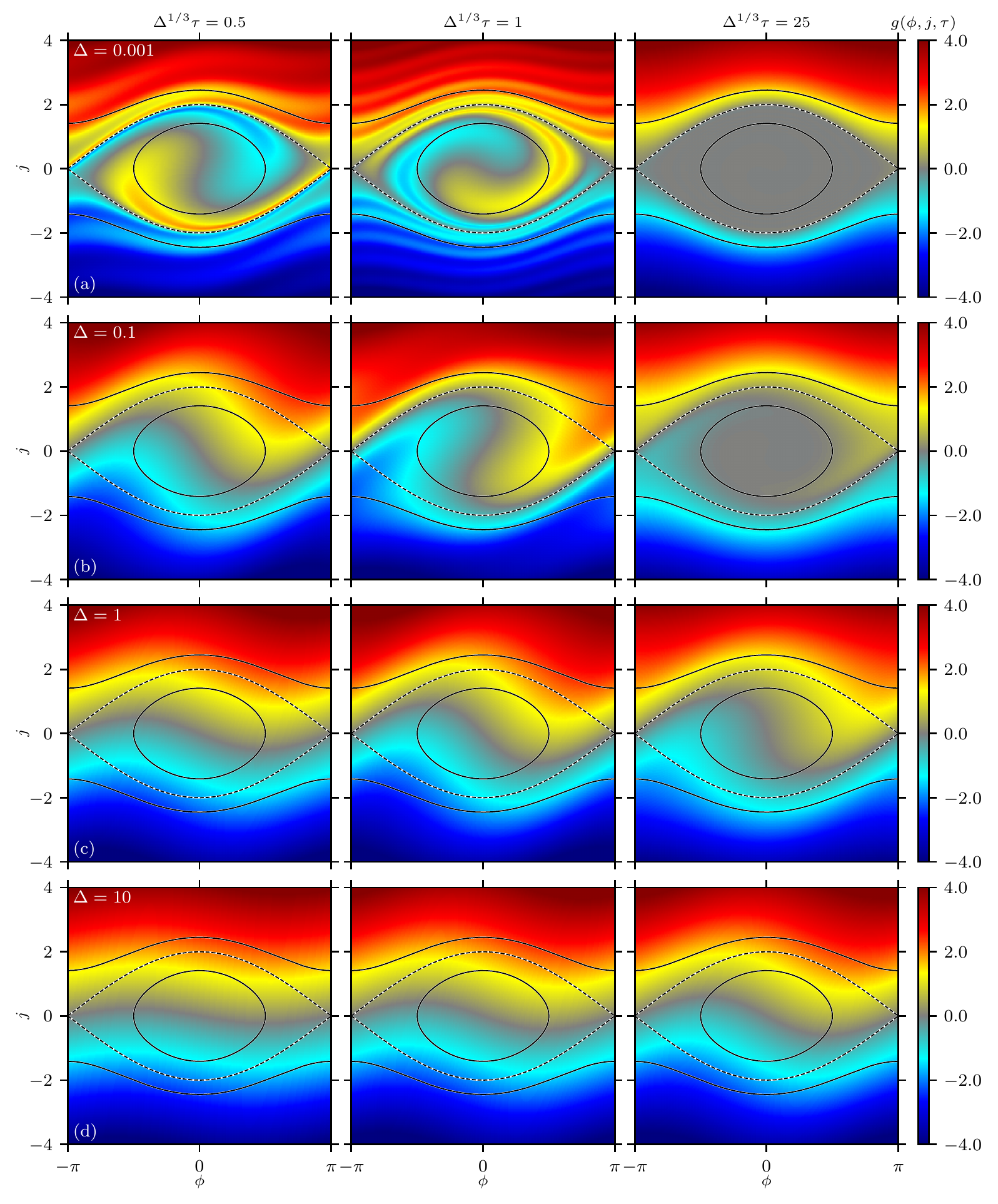}
\caption{Contours of $g(\phi, j, \tau)$ (related to $f$ through equation \eqref{eqn:dimensionless_DF}), computed by solving the kinetic equation (\ref{eqn:kinetic_equation_dimensionless}) numerically with the initial condition shown in Figure \ref{fig:dimensionless_DF}. In rows (a)--(d) we show the time evolution of the solution for $\Delta = 0.001,\, 0.1,\, 1,\, 10$ respectively. The right column represents the steady-state solution.}
\label{fig:slices}
\end{figure*}

In Figure \ref{fig:slices} we plot contours of the numerical solution $g(\phi, j, \tau)$ for $\Delta = 0.001,\, 0.1, \,1,\, 10$ in rows (a)--(d) respectively.  In each row, from left to right we plot the solution
at different `times' $\Delta^{1/3} \tau$. 

First consider row (a), which is in the limit of very weak diffusion ($\Delta =0.001$). Physically, we expect that in this limit the DF will \textit{phase mix} around and within the resonant island.
The reason for phase mixing is that in the absence of diffusion ($\Delta = 0$), the trajectories of individual particles trace contours of constant $\mathcal{H}$ in the $(\phi, j)$ plane; and since adjacent contours correspond to slightly different libration/circulation periods, the initial DF gets sheared out along these contours, {producing ever finer-scale structures in the phase space. If $\Delta$ were exactly zero, and in the absence of any coarse-graining, 
this process would continue indefinitely.
However, for our very small but finite value of $\Delta = 0.001$, diffusion is able to wipe out the finest-grained features, without needing to invoke any coarse-graining\footnote{In reality all systems have a finite number of particles, and hence a non-zero $\Delta$, regardless of how much one tries to suppress diffusion; there is no such thing as a perfectly collisionless system \citep{Silva2019}.}.
We see that by the third column, $g\approx 0$ inside the separatrix, and is smeared almost evenly
on contours of constant $\mathcal{H}$ outside of the separatrix. The DF has reached steady state.}

Now consider {row} (d), which corresponds to the opposite limit of very strong diffusion, represented here by $\Delta = 10$.  In this case, the resonance has little effect on $g$ at any time. This again is as expected since the initial linear DF (\ref{eqn:linear_DF}) is annihilated by the collision operator $C[f]$ (equation (\ref{eqn:collision_operator})). In other words, wherever the bar perturbation induces some curvature in the DF, strong diffusion immediately acts to remove it and restore the linear initial condition.

The cases $\Delta = 0.1$ and $\Delta =1$ (rows (b) and (c)) are intermediate between these two extremes.  
In the following subsections we unpick further the different features of these solutions and explain how their behavior depends on $\Delta$.
A reader who is satisfied with the basic picture shown in Figure \ref{fig:slices} can skip these details, and go directly to the more astrophysically interesting \S\ref{sec:Dynamical_friction}.

\subsubsection{Skew-symmetry}
\label{sec:skew_symmetry}

In every panel of Figure \ref{fig:slices} we notice the following skew-symmetry property:
\begin{equation}
    g(\phi, j, \tau) = -g(-\phi, -j, \tau).
    \label{eqn:skew}
\end{equation}
This follows easily from the symmetry of equation \eqref{eqn:kinetic_equation_dimensionless} under the replacements $\phi \to -\phi$ and $j \to -j$, and the corresponding
antisymmetry of the initial condition $g_\mathrm{init}(j) = j$.

\subsubsection{Steady state}
\label{sec:steady_state}

In each row (a)--(d) of Figure \ref{fig:slices} we find that by the third column the solution has approximately reached its steady state, which
reach reflects a balance between the secular bar torque (which wants to churn the DF around the island) and diffusion (which wants to restore the linear initial condition). The fact that a steady state is possible when our externally-imposed 
diffusion is constantly injecting energy into the system is a consequence of (i) the fact that the boundary condition is set to match the unperturbed DF at $\pm \infty$, which act as a particle source/sink.

How long does it take to reach the steady state?
 For $\Delta \ll 1$, where the bar resonance dominates, the typical steady-state timescale is a few libration periods, say
\begin{equation}
    t_\mathrm{steady} \sim 3 \, t_\mathrm{lib} \sim 3 \Delta \, t_\mathrm{diff}, \,\,\,\,\,\,\,\,\,\,\,\, (\Delta \ll 1)
    \label{sec:steady_state_timescale_weak}
\end{equation}
which corresponds to $\Delta^{1/3} \tau_\mathrm{steady} \lesssim 20$  --- see equation (\ref{eqn:dimensionless_time}).   Thus in this weakly-diffusive limit we typically have $t_\mathrm{lib} \ll t_\mathrm{steady} \ll t_\mathrm{diff}$.
Meanwhile in the diffusion-dominated regime $\Delta \gg 1$, we show in Appendix \ref{sec:strong_scattering} that
\begin{equation}
t_\mathrm{steady} \sim \frac{t_\mathrm{lib}}{\Delta^{1/3}}  \sim \Delta^{2/3} t_\mathrm{diff},\,\,\,\,\,\,\,\,\,\,\,\, (\Delta \gg 1)
\label{sec:steady_state_timescale_strong}
\end{equation}
i.e. $\Delta^{1/3} \tau_\mathrm{steady} \sim 2$.
Hence in this case we have the hierarchy 
$t_\mathrm{diff} \ll t_\mathrm{steady} \ll t_\mathrm{lib}$.
In general, for a given bar strength, stronger diffusion always leads to a more rapidly achieved steady state.

\subsubsection{Angle-averaged distribution}
\label{sec:angle_averaged}

\begin{figure*}[t]
\centering
\includegraphics[width=0.99\linewidth,trim={0.15cm 0cm 1.0cm 0cm},clip]{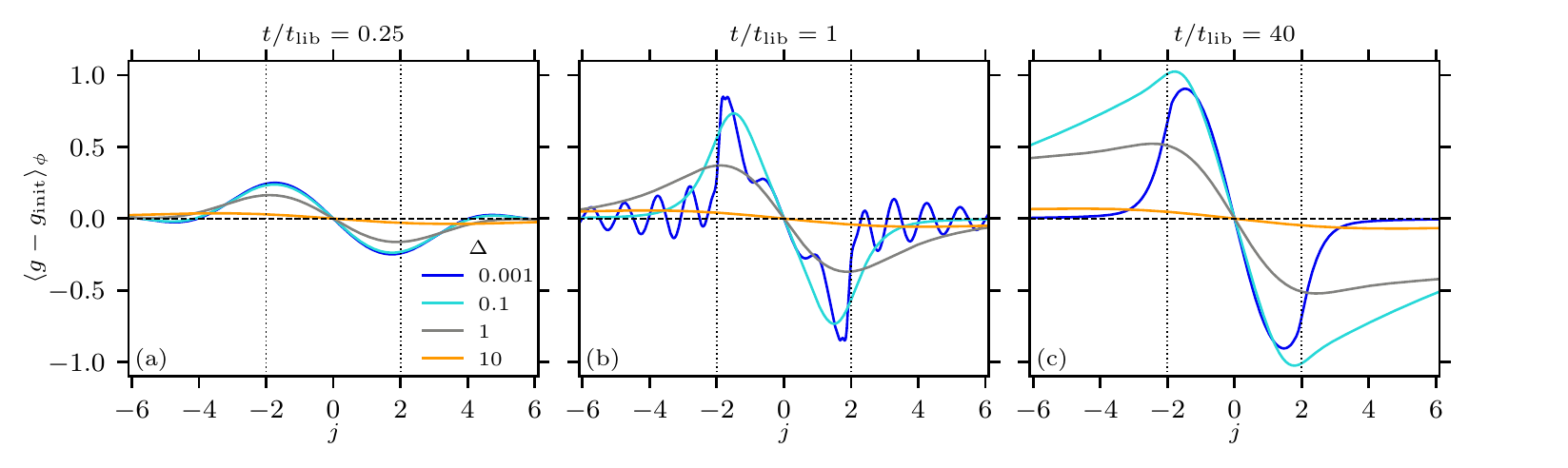}
\caption{Time evolution of the slow-angle-averaged perturbation to the DF,
$\langle g - g_{\rm init} \rangle_\phi$ (see equation \ref{eqn:phi_averaged_g}), for the same values of $\Delta$ as used in Figure \ref{fig:slices}. The right hand panel represents the steady-state solution. The vertical dotted lines show the maximum extent of the separatrix, $j = \pm 2$.}
\label{fig:phi_averaged_g}
\end{figure*}

It is instructive to average the solution $f(\phi, j, \tau)$ over the slow angle $\phi$ and investigate the resulting DF 
\begin{equation}
    \langle f \rangle_\phi \equiv \frac{1}{2\pi} \int_{-\pi}^\pi \md \phi \, f(\phi, j, \tau).
\end{equation}
In Figure \ref{fig:phi_averaged_g} we plot the corresponding auxiliary DF (see equation (\ref{eqn:dimensionless_DF})):
\begin{equation}
    \langle g - g_\mathrm{init} \rangle_\phi=   \frac{\langle f \rangle_\phi - f_0(0) }{\alpha} - j,
    \label{eqn:phi_averaged_g}
\end{equation}
as a function of $j$ for the same $\Delta$ values as in Figure \ref{fig:slices}.
In panels (a)--(c) we show this quantity for times $t/t_\mathrm{lib} = 0.25, 1, 40$ respectively; in particular, we know from \S\ref{sec:steady_state} that panel (c) always corresponds to the steady state. 
The vertical dotted lines in these panels are at $j=\pm 2$, which marks the maximum extent of the separatrix in the $(\phi, j)$ plane (Figures \ref{fig:dimensionless_DF}, \ref{fig:slices}).
We notice immediately that $\langle g - g_\mathrm{init} \rangle_\phi$
is an odd function of $j$, which follows from the skew-symmetry (\ref{eqn:skew}), and that in the steady-state
(panel (c)) it exhibits a single maximum and a single minimum.
For very small $\Delta$, the amplitude of these extrema actually grows with $\Delta$, peaking around $\Delta \sim 0.01$,
before decaying as $\Delta$ is increased further. 
Furthermore, in the very small $\Delta$ limit the location of the extrema is $\vert j\vert  \approx 1.5$, which is inside the maximum extent of the separatrix.
Increasing $\Delta$  broadens the distribution and shifts the locations of the extrema to larger $\vert j\vert$. For instance, for $\Delta = 1$ the amplitude of the $\langle g - g_\mathrm{init} \rangle_\phi$ curve 
is around half of what it was for $\Delta = 0.001$ while the locations of its extrema have shifted outside the separatrix to $\vert j \vert \approx 2.5$.
In the strong-diffusion regime $\Delta \gg 1$, the distribution becomes very broad and its amplitude decreases dramatically.

{We note that $\langle g - g_\mathrm{init}\rangle_\phi$
does not appear to decay towards zero at large $\vert j \vert$.
This is a well-known artefact of choosing the background DF to be linear and the collision operator to be of diffusion form (\citealt{Duarte2019-pl}; if we chose a Krook operator instead, we would not have the same issue).
This is the reason we force a boundary condition at the edge of the grid in $j$ (see \S\ref{sec:numerical_method}). We take the grid in $j$ sufficiently large that the sharp boundary does not feed back on the near-resonant dynamics. This issue also does not affect the bar-halo friction calculations in \S\ref{sec:Dynamical_friction}, since these are sensitive only to the angle-dependent part of the DF, which does decay for large $\vert j \vert$. 
}

\begin{figure*}[t]
\centering
\includegraphics[width=0.96\linewidth,trim={0.15cm 0cm 1.0cm 0cm},clip]{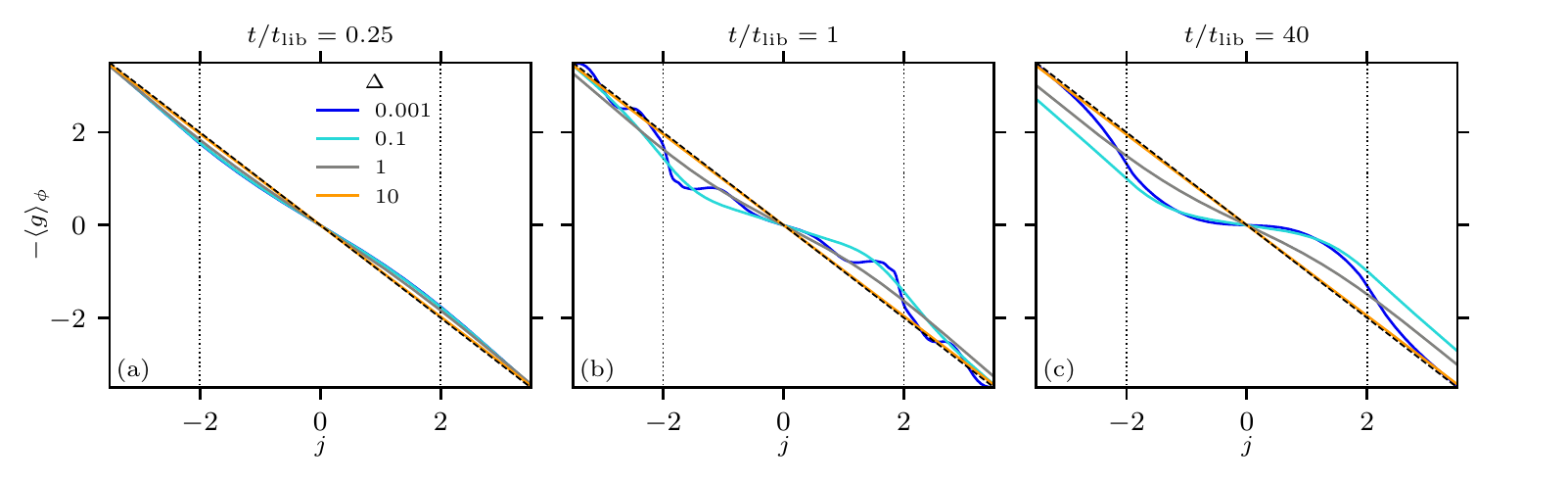}
\caption{As in Figure \ref{fig:phi_averaged_g}, except we show the evolution of
$-\langle g \rangle_\phi$, which is the dimensionless representation of a linearly-decreasing angle-averaged DF.}
\label{fig:phi_averaged_DF}
\end{figure*}

{To make clearer the evolution of the angle-averaged DF, we present Figure \ref{fig:phi_averaged_DF}, in which we use the same data as in Figure \ref{fig:phi_averaged_g} but this time we simply plot $-\langle g \rangle_\phi$.
This is a dimensionless representation of an initially linearly-decreasing, angle-averaged DF.
We see that for very small $\Delta \ll 1$ and late times (panel (c), blue lines), the angle-averaged DF has \textit{flattened} in the vicinity of the resonance.  Increasing $\Delta$ broadens the resonance, allowing it to affect larger $j$ values, but also weakens it, so that for very large $\Delta \gg 1$ (orange lines) the DF barely evolves at all.
We will quantify the broadening and weakening in Figure \ref{fig:Im_f1} below (see \S\ref{sec:asymmetry}).}

In Figure \ref{fig:Max_Values}a we plot the maximum value of $\langle g - g_\mathrm{init} \rangle_\phi$ in steady state as a function of $\Delta$.
We see that this quantity is roughly constant at small $\Delta \ll 1$, and of order unity in amplitude.
For large $\Delta \gg 1$
it decays like $\propto \Delta^{-1}$, which can also be shown analytically (\citealt{Duarte2019-mi}, equation (13)). It makes physical sense that the maximum value of $\langle g - g_\mathrm{init}\rangle_\phi \to 0$ as $\Delta \to \infty$, since for infinitely strong diffusion the DF should never evolve from its linear initial condition.

\subsubsection{Asymmetry in slow angle}
\label{sec:asymmetry}
\begin{figure}
\centering
\includegraphics[width=0.49\textwidth]{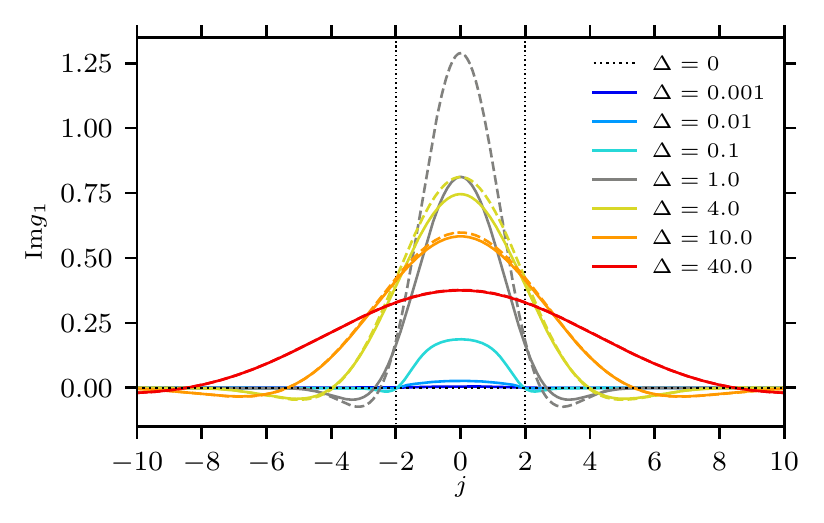}
\caption{Plot of $\mathrm{Im}\,g_1(j, \tau) \equiv \alpha^{-1} \mathrm{Im} \, f_1(j, \tau)$  --- see equation (\ref{eqn:f1}) --- as a  function of $j$ 
in the steady state for $\tau \to \infty$. Solid lines show the `exact' result from the numerical solution of equation \eqref{eqn:kinetic_equation_dimensionless}. Dashed lines show the analytical prediction from equations (\ref{eqn:Img1})--(\ref{eqn:resonance_function}) for $\Delta \geq 1$. The vertical dotted lines again show the maximum extent of the separatrix, $j = \pm 2$.}
\label{fig:Im_f1}
\end{figure}
Finally, when computing the dynamical friction torque on the bar in \S\ref{sec:Dynamical_friction} it will turn out that the key quantity we must extract from
our kinetic theory is Im $f_1(j, \tau)$ where $f_1$ is the first Fourier coefficient of the DF:
\begin{equation}
    f_1(j, \tau) \equiv \frac{1}{2\pi}\int_{-\pi}^{\pi} \md \phi \, f(\phi, j, \tau) \exp(-i\phi).
    \label{eqn:f1}
\end{equation}  
This quantity is obviously a measure of the asymmetry of the (slow-)angular distribution of particles; when there is no such asymmetry, there can be no frictional torque. 

In Figure \ref{fig:Im_f1} we plot the dimensionless quantity $\mathrm{Im} \, g_1 = \alpha^{-1} \mathrm{Im} \, f_1$ as a function of $j$ in the steady state $\tau\to \infty$, for various $\Delta$. 
We see from Figure \ref{fig:Im_f1} that $\mathrm{Im} \, g_1$ is always even in $j$, which follows from the skew-symmetry (\ref{eqn:skew}).  Its amplitude is very small in the weak-diffusion regime $\Delta \ll 1$, reflecting the (approximate) symmetry in $\phi$ of the (nearly) phase-mixed steady-state --- see Figure \ref{fig:slices}a,b. Indeed 
we know that in the perfectly collisionless regime
\begin{equation}
    \mathrm{Im}\, g_1(j, \tau \to \infty) =0, \,\,\,\,\,\,\,\,\,\,\,\, (\Delta =0)
    \label{eqn:Img1_TW84}
\end{equation}
However, once the diffusion strength $\Delta$ is increased beyond $\sim 0.1$ the peak value of $\mathrm{Im}\,g_1$ grows significantly,  then begins to decay with $\Delta$ for $\Delta \gg 1$. 
The dashed curves we show for $\Delta \geq 1$ in this plot
correspond to the approximate analytic solution valid for $\Delta \gg 1$ derived in \S\ref{sec:strong_scattering}, namely
\begin{equation}
    \mathrm{Im}\, g_1(j, \tau \to \infty) \approx \frac{\pi}{2} \mathcal{R}_\Delta(j), \,\,\,\,\,\,\,\,\,\,\,\, (\Delta \gg 1)
    \label{eqn:Img1}
\end{equation}
where 
\begin{equation}
    \mathcal{R}_\Delta(j) \equiv \frac{1}{\pi \Delta^{1/3}} \int_0^\infty \md y \, \exp\left( -\frac{y^3}{3}\right)\cos \left( \frac{j y}{\Delta^{1/3}}\right).
    \label{eqn:resonance_function}
\end{equation}
It is not hard to show that $\mathcal{R}(j)$ has the properties of a collisionally-broadened resonance function \citep{Duarte2019-mi}, namely
$\int^\infty_{-\infty} \md j \, \mathcal{R}_\Delta(j) =1 $ and $\lim_{\Delta \to 0} \mathcal{R}_\Delta(j) = \delta(j)$. 
We see from Figure \ref{fig:Im_f1} that the approximation (\ref{eqn:Img1}) is accurate to within several percent or better for $\Delta \gtrsim 4$.
It is also apparent from Figure \ref{fig:Im_f1} that the width of the $\mathrm{Im}\, g_1(j, \tau\to \infty)$ curve increases monotonically with $\Delta$.
Indeed, in the $\Delta \gg 1$ limit we know from equations (\ref{eqn:Img1})-(\ref{eqn:resonance_function}) that this width grows like $\propto \Delta^{1/3}$, and that the total area under the $\mathrm{Im}\, g_1(j, \tau \to \infty)$ curve is constant.

The reason for this lack of $\Delta$-dependence at large $\Delta \gg 1$ is that 
strong diffusion essentially renders the bar-halo interaction linear (see the Discussion).
Unlike in the nonlinear phase, the linear dynamics is insensitive to time delay effects that arise from successive integrations of the kinetic equation within a perturbative approach \citep{Berk1996-wp}.

\begin{figure}
\centering
\includegraphics[width=0.49\textwidth]{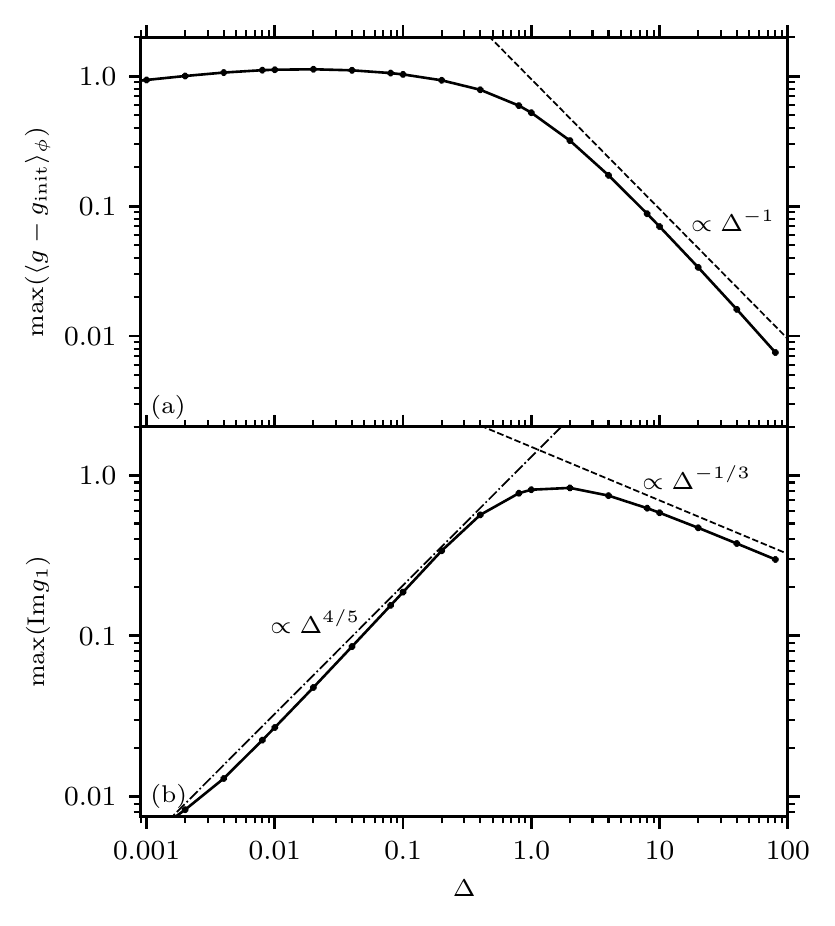}
\caption{Plots of the maximum steady-state values of (a) $\langle g - g_\mathrm{init} \rangle_\phi$ (see Figure \ref{fig:phi_averaged_g}), and (b) $\mathrm{Im} \, g_1$ (see Figure \ref{fig:Im_f1}) as functions of $\Delta$.}
\label{fig:Max_Values}
\end{figure}

Lastly, in Figure \ref{fig:Max_Values}b we extract the steady-state value of $\mathrm{max}( \mathrm{Im}\, g_1)$ and plot it as a function of $\Delta$.
For small $\Delta \ll 1$ we have roughly $\mathrm{max}( \mathrm{Im} \, g_1) \propto \Delta^{4/5}$.  This is an empirical scaling which is hard to explain mathematically (see \S\ref{sec:weak_scattering}). {Intuitively, it} stems from the fact that weak diffusion `fills in' the DF near the inner edge of the separatrix, where we would have $g=0$ were there no diffusion at all;
The amplitude of $g$ in the filled-in `strip' is $\mathcal{O}(1)$, while the strip's thickness grows monotonically with $\Delta$.
For large $\Delta \gg 1$ we see that $\mathrm{max}( \mathrm{Im} \, g_1) \propto \Delta^{-1/3}$, which agrees with (\ref{eqn:Img1})--(\ref{eqn:resonance_function}).  The transition between these two regimes occurs around $\Delta \sim 2$, where $\mathrm{max}( \mathrm{Im} \, g_1) \sim 1$.



\section{Bar-dark matter halo dynamical friction}
\label{sec:Dynamical_friction}

As a galactic bar ages, it transfers angular momentum to its host dark matter halo and consequently its rotation rate slows.
The mechanism responsible is dynamical friction: the bar produces a perturbation in the dark matter distribution function $f$ and hence in the dark matter density, which then back-reacts to produce a torque $\mathcal{T}$ on the bar, draining its angular momentum. 
Problems of bar-halo coupling --- and more generally the problem of angular momentum transfer between a massive perturber and a distribution of particles --- 
have been the focus of many classic studies in galactic dynamics (e.g. \citealt{Lynden-Bell1972-ve,Tremaine1984-wt,Weinberg1985-en,Debattista1998-nv,Athanassoula2003-xj}) and continue to inspire modern research \citep{Kaur2018-wp,Chiba2020-th,Collier2021-nf,Banik2021-ug,Lieb2021-tv,Chiba2022-qt,Kaur2022-tb,Dootson2022-cu}.  They are also strongly analogous to wave-particle interaction problems in plasma (see the Discussion).

One can write down a rather general formula for the dynamical friction torque $\mathcal{T}$ on the bar as follows.
Like we have throughout this paper, let the host dark matter halo be spherical and let the bar rotate around the $z$-axis and have potential $\delta \Phi(\bm{x},t)$.
From Hamilton's equation the $z$-component of the specific torque felt by an individual dark matter particle due to the bar is
\begin{equation}
    \frac{\md L_z}{\md t} = -\frac{\partial \delta \Phi}{\partial \theta_\varphi}.
\end{equation}
Let the DF of dark matter particles be $f$, {normalized such that $\int \md \btheta \md \bJ f =1$.}  Then by Newton's third law and using $\md \bm{x} \md \bm{v} = \md \btheta \md \bJ$,
the total torque on the bar (divided by the mass of the halo) due to the dark matter particles is equal to 
\begin{equation}
\mathcal{T}(t) =  \int \md \btheta \md \bJ \, 
f(\btheta, \bJ,t) \frac{\partial \delta \Phi(\btheta, \bJ, t)}{\partial \theta_\varphi}.
\label{eqn:torque_general}
\end{equation}
The challenge is to compute $f$ for a given perturbation $\delta \Phi$, and then perform the integral \eqref{eqn:torque_general}. 

\subsection{Linear theory}
\label{sec:Dynamical_friction_linear}

\cite{Lynden-Bell1972-ve} (hereafter LBK) and \cite{Weinberg2004-ss} compute $f$ using the linearized collisionless Boltzmann (Vlasov) equation,
ignoring the self-gravity of the perturbation to the dark matter distribution.  Fourier expanding the potential as $\delta \Phi = \sum_{\bm{N}} \delta \Phi_{\bm{N}}(\bJ,t) \exp(i\bm{N}\bcdot \bm{\theta})$ and similarly for the DF $f$, the linear phase space response induced by the bar is
\begin{eqnarray}
    f_{\bm{N}}(\bJ, t) = i \bm{N} \bcdot \frac{\partial f_0}{\partial \bJ}
   \int_0^t \md t' \delta\Phi_{\bm{N}} (\bm{J}, t') \me^{-i\bm{N}\bcdot \bm{\Omega}(t-t')}, \nn \\
\end{eqnarray}
where $f_0(\bJ)$ is the unperturbed DF.
Then putting $\delta\Phi_{\bm{N}} (\bm{J}, t) = \delta\Phi_{\bm{N}} (\bm{J}) \exp(-i N_\varphi \pattern t)$ and
performing the integral over $t'$, one finds a `linear torque' $ \mathcal{T}^\mathrm{lin} = \sum_{\bm{N}}  \mathcal{T}^\mathrm{lin}_{\bm{N}}$, where the contribution from resonance $\bm{N}$ is:
\begin{eqnarray}
          \mathcal{T}^\mathrm{lin}_{\bm{N}}(t) \equiv (2\pi)^3  N_\varphi \int && \md \bJ  \vert \delta \Phi_{\bm{N}}  \vert^2 \bm{N}\bcdot \frac{\partial f_0}{\partial \bJ} \nn \\ && \times \frac{\sin[(\bm{N}\bcdot \bm{\Omega} - N_\varphi \pattern)t]}{\bm{N}\bcdot \bm{\Omega} - N_\varphi \pattern}.
    \label{eqn:Torque_linear}
\end{eqnarray}
Taking the
time-asymptotic limit $t\to \infty$ one arrives at the classic `LBK torque formula':
\begin{eqnarray}
        \mathcal{T}^\mathrm{LBK}_{\bm{N}} \equiv (2\pi)^3  N_\varphi \int && \md \bJ  \vert \delta \Phi_{\bm{N}}  \vert^2 \bm{N}\bcdot \frac{\partial f_0}{\partial \bJ}  \nn \\ & & \times \pi \delta \left(\bm{N}\bcdot \bm{\Omega} - N_\varphi \pattern \right),
    \label{eqn:LBK}
\end{eqnarray}
which predicts that angular momentum is transferred to and from the dark matter halo exclusively at resonances.
Importantly, for many {realistic} dark matter DFs $f_0(\bJ)$, the LBK torque is finite and negative, implying a long-term transfer of angular momentum away from the bar and hence a decay in its pattern speed. In practice the time-asymptotic limit may not be valid since the torque {takes some time to converge, by which time the pattern speed may have changed significantly
 \citep{Weinberg2004-ss}.}
Nevertheless, the LBK formula (\ref{eqn:LBK}) is a good benchmark against which we can compare the magnitude of the torque arising from more sophisticated calculations.

\subsection{Nonlinear theory}

Since they employ linear theory, LBK and \cite{Weinberg2004-ss} do not account for the inevitable nonlinear particle trapping that occurs sufficiently close to each resonance.
Recognizing this shortcoming, \citet{Tremaine1984-wt} --- hereafter TW84 ---  and \cite{Chiba2022-qt}
{include} the particle trapping effect.  To do this they first convert to slow-fast angle-action variables $(\bm{\theta}',\bm{J}')$ \textit{around each resonance} $\bm{N}$ as in \S\ref{sec:Single_particle}.  In these variables
the total `nonlinear torque' on the bar $\mathcal{T}^\mathrm{nl}$ can be written as a sum over $\bm{N}$ of contributions
\begin{equation}
\mathcal{T}^\mathrm{nl}_{\bm N}(t) = \int \md \bm{\theta}' \md \bm{J}' \, 
f(\bm{\theta}', \bm{J}', t) 
N_\varphi \frac{\partial \delta \Phi(\btheta', \bJ')}{\partial \theta_\mathrm{s}}.
\label{eqn:Torque_slowfast}
\end{equation}
If we now Fourier expand the potential $\delta \Phi$ in slow angle space as in equation (\ref{eqn:Hamiltonian_slowfastangles}), and also expand the DF as $f = \sum_{\bm{k}} f_{\bm{k}}(\bJ',t) \exp(i \bm{k}\bcdot \bm{\theta}')$, then \eqref{eqn:Torque_slowfast} reads
\begin{equation}
\mathcal{T}^\mathrm{nl}_{\bm N}(t) = - i(2\pi)^3 N_\varphi 
\sum_{\bm{k}} k_\mathrm{s}
\int 
\md \bm{J}' \, 
f_{\bm{k}}(\bm{J}', t) 
   \Psi^*_{\bm{k}}(\bm{J}'),
   \label{eqn:friction_slowfast}
\end{equation}
where one must be careful to properly divide phase space into non-overlapping sub-volumes that are dominated by individual resonances (see \S4.5 of \citealt{Chiba2022-qt} for a discussion).
%
Since we always choose our initial $f$ to be independent of angles (i.e. dependent only on $\bJ'$), we know from \S\ref{sec:Single_particle} that there will be no fast-angle dependence in $f$ at later times.  This means that $f_{\bm{k}}= \delta_{k_1}^0 \delta_{k_2}^0 f_{(0,0,k_\mathrm{s})}$, so
\begin{eqnarray}
\mathcal{T}^\mathrm{nl}_{\bm N}(t) =
    2(2\pi)^3 N_\varphi 
\sum_{k>0} k
\int 
\md \bm{J}' 
\, \mathrm{Im} \left[
f_{k}(\bm{J}', t) 
   \Psi^*_{k}(\bm{J}') \right], \nn \\ 
    \label{eqn:nonlinear_torque}
\end{eqnarray}
where we have used the shorthand $f_{(0,0,k)} = f_k$, and similarly for $\Psi_k$.
At this stage, \cite{Chiba2022-qt} calculate the torque $\mathcal{T}^\mathrm{nl}_{\bm N}(t)$ by substituting for $f_k$ the 
collisionless DF, i.e. the solution to the kinetic equation \eqref{eqn:kinetic_equation_dimensionless} with $\Delta =0$.
In the time-asymptotic limit they arrive at the classic {TW84} result
\begin{equation}
    \mathcal{T}^\mathrm{TW84}_{\bm N} = 0.
    \label{eqn:TW84}
\end{equation}
This zero torque result is a consequence of the symmetry of the dark matter density distribution that arises when 
the DF completely phase mixes within and around the resonant island.
Simply put, in the fully phase-mixed state there are the same number of particles `pushing' on the bar as `pulling' on it --- see Figures 4, 7 and 16 
of \cite{Chiba2022-qt} for illustration.
\\
\\
The linear and nonlinear torque calculations pursued by the above authors (LBK; \citealt{Weinberg2004-ss}; TW84; \citealt{Chiba2022-qt}) and similar efforts by others
\citep{Weinberg1989-cj,Banik2021-ug,Banik2022-rj,Kaur2018-wp,Kaur2022-tb}
were all \textit{collisionless}: their DF $f$ only responded to the 
smooth combined potential of the underlying equilibrium galaxy $\Phi_0$ and the rigidly rotating perturbation $\delta \Phi$, in effect setting $\Delta = 0$.
The only related (semi-)analytical study we know of to have included
the effects of diffusion ($\Delta > 0$) in this problem is \cite{Weinberg2007-bv},
but their focus was on the evolution of the bar pattern speed and the associated rate at which resonances sweep through phase space {(see \S\ref{sec:WeinbergSellwood})}.
In the remainder of this section we 
will investigate how finite $\Delta$ affects the dynamical friction torque on galactic bars, 
focusing on the corotation resonance $\bm{N} = (0,2,2)$ identified as dominant by \cite{Chiba2022-qt}.

\subsection{The corotation torque density}
\label{sec:corotation_torque_density}

\begin{figure}
\includegraphics[width=0.49\textwidth]{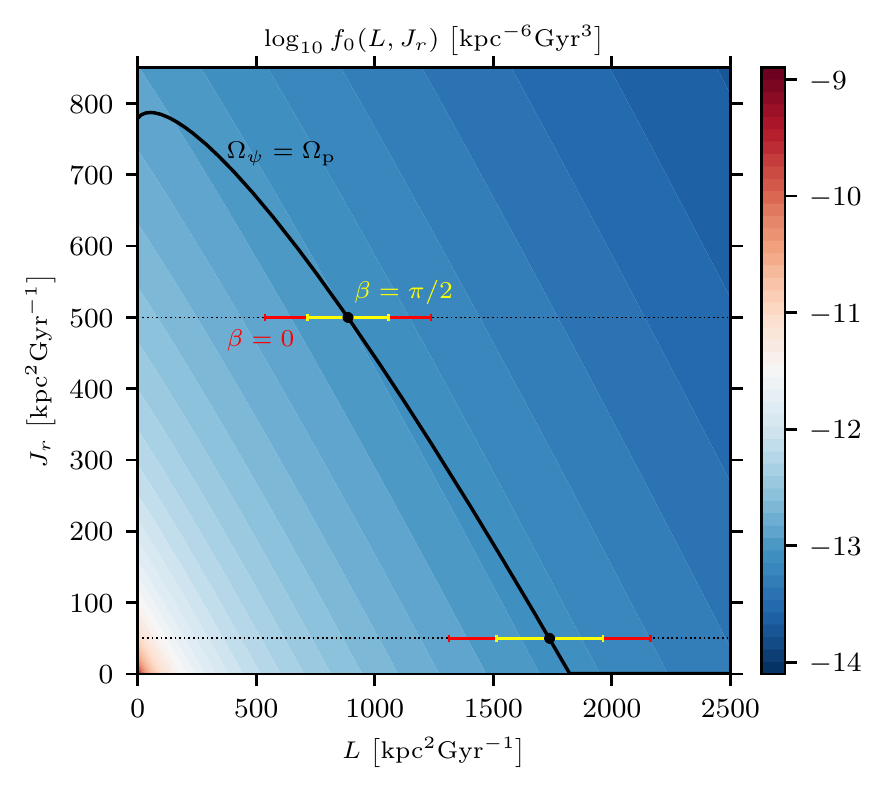}
\caption{Contours of the isotropic Hernquist DF, $\log_{10} [f_0/(\mathrm{kpc}^{-6}\mathrm{Gyr}^3)]$, as a function of angular momentum $L$ and radial action $J_r$ for arbitrary incination $\beta$.
The solid black line shows the corotation resonance, $\Omega_\psi=\pattern = 35 \mathrm{km}\,\mathrm{s}^{-1}\mathrm{kpc}^{-1}$, 
which is also independent of $\beta$.  We choose two particular locations on the resonant line, shown with black dots, corresponding to $J_r = (500, 50)$ kpc$^2$ Gyr$^{-1}$. Around each black dot we show with red (yellow) `error bars' the maximum extent of the separatrix in $L$ for $\beta = 0$ ($\beta = \pi/2$); these bars therefore mark out the region $j \in (-2,2)$.}
\label{fig:Resonant_Line}
\end{figure}
We want to calculate the torque on the bar using the nonlinear torque equation
(\ref{eqn:nonlinear_torque}), and substituting for $f$ our solution to the kinetic equation discussed in \S\ref{sec:solution}, for different values of $\Delta$.
Following \cite{Chiba2022-qt} we use as our dark matter halo model the Hernquist sphere (\ref{eqn:Hernquist_potential}), and we take our unperturbed dark matter DF $f_0(\bJ)$ 
to be the isotropic Hernquist DF \citep{Hernquist1990-oe}.  In Figure \ref{fig:Resonant_Line} we show colored contours of $f_0(L, J_r)$ at fixed (arbitrary) inclination 
\begin{equation}
    \beta \equiv \arccos(L_z/L).
\end{equation}
For our bar potential we use the model (\ref{eqn:bar_explicit})--(\ref{eqn:bar_details}), with the choices of numerical parameters given below equation (\ref{eqn:bar_details}).
We will focus only on the corotation resonance $\bm{N} = (0,2,2)$, so the implicit equation for the resonant line in phase space is 
\begin{equation}
    \Omega_\psi(L, J_r) = \Omega_\mathrm{p},
\end{equation}
with $\Omega_\mathrm{p} = 35 \, \mathrm{km} \, \mathrm{s}^{-1}\mathrm{kpc}^{-1} = 35.8$ Gyr$^{-1}$.
We plot this resonant condition with a solid black line in Figure \ref{fig:Resonant_Line}. We also choose two particular resonant locations, shown with black dots, corresponding to $J_r = 500$ kpc$^2$ Gyr$^{-1}$ and 
 $J_r = 50$ kpc$^2$ Gyr$^{-1}$ respectively, and will refer to these momentarily.

Next, given the choice of $\bm{N}$ it follows from (\ref{eqn:fast_actions})--(\ref{eqn:slow_action}) that
\begin{eqnarray}
&&\bJ_\mathrm{f} = \left(J_r, \, L(1-\cos \beta) \right),
\label{eqn:Corotation_fast_actions}
\\
  &&  J_\mathrm{s} = \frac{L \, \cos \beta}{2},
  \label{eqn:Corotation_slow_action}
\end{eqnarray}
Thus at fixed inclination, $L$ is (proportional to) the slow action, while $J_r$ is a fast action.
Suppose the bar is initially absent and the DF is $f_0(\bJ)$, and then at $t=0$ we suddenly turn on the bar perturbation.
Particles that were initially on zero inclination orbits in the background spherical potential
($\beta = 0$) will remain at zero inclination even under the finite bar perturbation, which follows from the conservation of the 
fast action $J_{\mathrm{f}2}$ (equation (\ref{eqn:Corotation_fast_actions})).
As a result, for these particles $L$ is a genuine slow action (up to a factor of $2$), meaning they move in the $(L, J_r)$ phase space only along the horizontal lines of constant $J_r$, for instance along one of the black dotted horizontal lines shown in Figure \ref{fig:Resonant_Line}.
Ignoring diffusion for now, 
particles that are initially sufficiently close to the resonance will be trapped by it and will oscillate back and forth across the solid black resonant line (librating orbits).
Particles that are somewhat further away from the resonance will also oscillate in $L$ but will not cross the resonant line (circulating orbits).

Before we proceed with our torque calculation, one conceptual hurdle must be overcome.
Namely, for a given diffusion coefficient $D$ the corotation torque $\mathcal{T}^\mathrm{nl}_{(0,2,2)}(t)$ involves 
contributions from a range of different $\Delta$ values.
To demonstrate this, in Figure \ref{fig:Resonant_Line}
we {show} the extent in $j$ of the librating orbits (i.e. the resonance width) for $\beta = 0$ with red bars; in other words these red bars correspond to $j \in (-2, 2)$.
We notice that the resonance width depends on the choice of $J_r$.
Similarly, with the yellow bars in Figure \ref{fig:Resonant_Line}
we show the resonance width for the same $J_r$ values but a different inclination, namely $\beta = \pi/2$, and again the width changes\footnote{There is a minor subtlety 
here: for orbits that are initially at nonzero inclination, the inclination precesses under the bar perturbation,
so the action-space evolution of an initially inclined particle cannot be fully captured in a 
single $\beta = $ constant diagram.  
Regardless, the width of the resonance clearly depends on $\beta$.}.
Moreover it is easy to show that the libration time $t_\mathrm{lib}$ depends on both $\beta$ and $J_r$.
It follows that
$\Delta$ (equation (\ref{eqn:dimensionless_diffusion})) is not a constant but rather a function of $(J_r, \beta)$.
The corotation torque $\mathcal{T}^\mathrm{nl}_{(0,2,2)}(t)$ involves an
integration over $L$, $J_r$ and $\cos\beta$ (see equation (\ref{eqn:nonlinear_torque})), and therefore has contributions from many different $\Delta$.
Since our purpose in this paper is to understand how the physics of resonances
  depends on $\Delta$ (rather than to provide the most accurate possible computation of the frictional torque) we choose to isolate a quantity that involves only a single value of $\Delta$.
This quantity is the \textit{corotation torque density} $ \mathcal{S}$, defined such that
\begin{equation}
\mathcal{T}^\mathrm{nl}_{(0,2,2)}(t) = \int 
\md J_r \, \md \cos\beta   \,   \mathcal{S}(t \vert J_r, \cos\beta).
\label{eqn:Total_Corotation_Torque}
\end{equation}
Thus $\mathcal{S}(t)$ measures the contribution to the total corotation torque from dark matter particles with radial actions $\in (J_r, J_r + \md J_r)$ and inclinations $\in (\cos\beta, \cos\beta + \md \cos \beta)$.  Importantly, for a fixed $J_r$ and $\cos \beta$, the torque density $\mathcal{S}$ has contributions from only a single value of $\Delta$,
which we {are therefore free to stipulate} by hand.  

The formula for $\mathcal{S}$ turns out to be
\begin{equation}
    \mathcal{S}(t \vert J_r, \cos\beta) = 2(2\pi)^3 \int \md L \,L  \Psi_1\,  \mathrm{Im} \, f_1,
    \label{eqn:corotation_torque_density}
\end{equation}
where (\citealt{Chiba2022-qt}, Appendix B):
\begin{equation}
    \Psi_1(\bJ) = \frac{(1+\cos\beta)^2}{8\pi} \int_0^\pi \md \theta_r \Phi_\mathrm{b}(r) \cos [2(\theta_\psi-\psi)].
    \label{eqn:corotation_potential_perturbation}
\end{equation}
For a given $\Delta$ value and choice of $(J_r, \cos\beta)$, we can compute $\mathrm{Im}\,f_1$ using the numerical solution to the kinetic equation \eqref{eqn:kinetic_equation_dimensionless} that we described in \S\ref{sec:KT_resonance}, setting the initial condition to be of the form (\ref{eqn:linear_DF}) by linearizing the isotropic Hernquist DF (shown in Figure \ref{fig:Resonant_Line}) around the resonance.
We can also calculate $\Psi_1(\bJ)$ (equation (\ref{eqn:corotation_potential_perturbation})) efficiently on a grid in $(L, J_r, \cos\beta)$ space using 
the standard mappings for $(\bm{x}, \bm{v}) \to (\bm{\theta}, \bm{J})$ for spherical potentials (e.g. \S3.5.2 of \citealt{Binney2008-ou}), and the `angular anomaly' trick from Appendix B of \cite{Rozier2019-ms}. As mentioned below equation (\ref{eqn:friction_slowfast}), when performing the integral in (\ref{eqn:corotation_torque_density}) one has to choose the maximum/minimum $L$ (or equivalently, $j$) values at which to cut it off, and this choice will affect the results, as we will see. 

\subsection{Results}
\label{sec:numerical_results}

{In this section we present the results of our corotation torque density calculations for various $\Delta$.
A reader uninterested in the details can skip to the summary in \S\ref{sec:friction_summary}.} 

\begin{figure*}
\centering
\includegraphics[width=\textwidth]{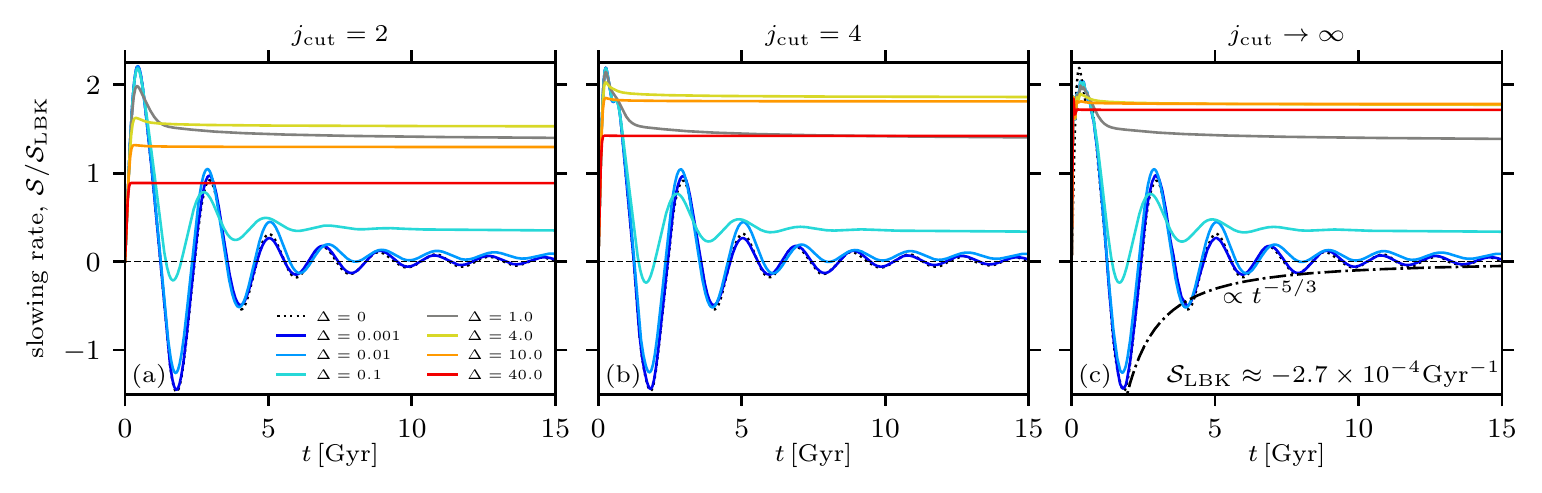}
\caption{Dimensionless slowing rate of the bar 
$\mathcal{S}(t) / \mathcal{S}^\mathrm{LBK}$ due to halo friction, specifically from dark matter particles with $\beta = 0$ and $J_r = 500$ kpc$^2$ Gyr$^{-1}$. The libration time is $t_\mathrm{lib} = 1.9$ Gyr and the LBK corotation torque density is $S_\mathrm{LBK} = - 2.7 \times 10^{-4} \mathrm{Gyr}^{-1}$.
Colored lines show results for different $\Delta > 0$, while the dotted line shows the collisioness ($\Delta = 0$) result.
In panels (a)--(c) we cut off the integral in (\ref{eqn:corotation_torque_density}) at $j_\mathrm{cut} = 2$, $4$ and $\infty$ respectively.}
\label{fig:Torque_time_1}
\end{figure*}

In Figure \ref{fig:Torque_time_1} we show the bar's dimensionless slowing rate
\begin{equation}
    \mathcal{S}(t) / \mathcal{S}_\mathrm{LBK},
    \label{eqn:torque_on_halo}
\end{equation}
where $\mathcal{S}_\mathrm{LBK}$ is the 
LBK corotation torque density (see equation (\ref{eqn:LBK})):
\begin{eqnarray}
    \mathcal{S}_\mathrm{LBK}(J_r, \cos\beta) = 4(2\pi)^3 \int && \md L \,L\, \vert \Psi_1 \vert^2 \frac{\partial f_0}{\partial L}  \nn \\ && \times \pi \delta [2(\Omega_\psi - \pattern)], 
    \label{eqn:LBK_density}
\end{eqnarray}
which is the result of computing the torque density using linear theory with $\Delta = 0$.
In each panel we 
fix the inclination $\beta = 0$ and radial action $J_r = 500$ kpc$^2$ Gyr$^{-1}$, so the resonance location corresponds to the upper black dot in Figure \ref{fig:Resonant_Line}. The libration timescale (equation (\ref{eqn:t_libration})) for this resonance location is $t_\mathrm{lib} = 1.9$ Gyr.
In each panel of Figure \ref{fig:Torque_time_1} we plot the slowing rate (\ref{eqn:torque_on_halo}) for various values of $\Delta \geq 0$. 
Note that $\mathcal{S}_\mathrm{LBK}$ is negative, so that a positive value of $ \mathcal{S}(t) / \mathcal{S}_\mathrm{LBK}$ means the bar feels a negative torque (slowing it down).
The difference between the panels lies in our choice of integration limits in equation (\ref{eqn:corotation_torque_density}):
we choose the limits to be $j=\pm j_\mathrm{cut}$ for $j_\mathrm{cut} = 2$, $4$ and $\infty$ in panels (a)--(c) respectively.\footnote{{Really the third value is $j_\mathrm{cut}=10$, but this 
is essentially indistinguishable from $j_\mathrm{cut} \to \infty$.
Also, when $-j_\mathrm{cut}$ corresponds to a negative $L$, we set the minimum $j$ to whichever value corresponds to $L=0$.}} Note that $\mathcal{S}_\mathrm{LBK}$ (equation (\ref{eqn:LBK_density})) does not depend on $j_\mathrm{cut}$.

Let us focus first on panel (a), which is for $j_\mathrm{cut} = 2$, meaning that the edges of our integration domain just graze the maximum extent of the separatrix in the $(\phi, j)$ plane.
In the collisionless case ($\Delta = 0$, black dotted line) we see that the slowing rate oscillates on the timescale $\sim t_\mathrm{lib}$.
This is the same behavior as found by \cite{Chiba2022-qt} --- indeed, the $\Delta = 0$ results shown here are very similar to the top panel of their Figure 11.\footnote{Note however that they plot the total corotation torque $-\mathcal{T}_{(0,2,2)}$ {(equation \eqref{eqn:Total_Corotation_Torque})}, whereas we only plot the corotation torque density {\eqref{eqn:corotation_torque_density}}.}
{The envelope of the $\Delta = 0$ curve decays approximately as $\propto (t/t_\mathrm{lib})^{-5/3}$ (shown explicitly in panel (c)), so 
in} the time-asymptotic limit for $\Delta = 0$ we recover the TW84 result (c.f. equation (\ref{eqn:TW84})):
\begin{equation}
    \mathcal{S}(t\to \infty) = 0, \,\,\,\,\,\,\,\,\,\,\,\,\,\,\, (\Delta =0),
    \label{eqn:S_TW84}
\end{equation}
which follows from the symmetry of the fully phase-mixed DF in slow angles, i.e. $\mathrm{Im}\, f_1(j, \tau\to \infty) = 0$ (Figure \ref{fig:Im_f1}).
However, as one deviates from $\Delta = 0$ to finite (but small) $\Delta > 0$, the behavior changes.
The steady state is reached sooner than it was in the collisionless case, following our expectations from \S\ref{sec:steady_state}. In addition, the 
slowing rate in the steady state is manifestly positive for all $\Delta > 0$.
This is a consequence of the fact that while phase-mixing attempts to abolish the asymmetry of the the slow-angle distribution, 
diffusion replenishes it (e.g.  Figure \ref{fig:slices}b) leading to a finite $\mathrm{Im} \, f_1$ (Figure \ref{fig:Im_f1}).
Indeed the slowing rate in Figure \ref{fig:Torque_time_1} grows with $\Delta$ until, for $\Delta \sim 1$, it is actually larger than the LBK value.  This growing trend continues up to around $\Delta \sim 10$, when the slowing rate starts to decay with increasing $\Delta$, albeit rather gradually; even by $\Delta = 40$ the steady-state slowing rate is barely smaller than the LBK prediction.

This story mostly repeats itself in panels (b) and (c), i.e. for $j_\mathrm{cut} = 4$ and $j_\mathrm{cut} \to \infty$ respectively, with some key differences.
First of all, the transient early time behavior of the slowing rate ($t < t_\mathrm{lib}$) is sensitive to the choice of $j_\mathrm{cut}$, especially for small $\Delta$,  as might be expected e.g. from the first two panels of Figure \ref{fig:slices}a.
Yet for $t \gtrsim t_\mathrm{lib}$ the $\mathcal{S}(t)$ curves for small $\Delta \lesssim 1$ in Figure \ref{fig:Torque_time_1} take a near-universal form independent of $j_\mathrm{cut}$. 
This is because for $\Delta \lesssim 1$ the steady-state value of $\mathrm{Im}\,f_1(j)$ is negligible outside of the separatrix region $j \in (-2, 2)$  (see Figure \ref{fig:Im_f1}).
Physically, under very weak diffusion the disturbances to the initial DF produced by the bar cannot propagate very far beyond the resonant island (as reflected in e.g. Figure \ref{fig:phi_averaged_g}c), so it is not possible to produce angular asymmetries at large $j$.
This renders the slowing rate insensitive to $j_\mathrm{cut}$ as long as one chooses a value larger than about $2$.
Essentially the same conclusion was drawn by \cite{Chiba2022-qt} when they noted that the great majority of the torque in the collisionless limit came from trapped orbits as opposed to circulating ones.
On the contrary, in the strong-diffusion limit $\Delta \gg 1$ the slowing rate \textit{is} sensitive to $j_\mathrm{cut}$, because disturbances are able to propagate much further. Mathematically, for $\Delta \gg 1$ the curve of
$\mathrm{Im}\,f_1(j, \tau \to \infty)$ becomes very broad in $j$, with a width
$\propto \Delta^{1/3}$ --- see \S\ref{sec:asymmetry}.

%
\begin{figure}
\centering
\includegraphics[width=0.49\textwidth]{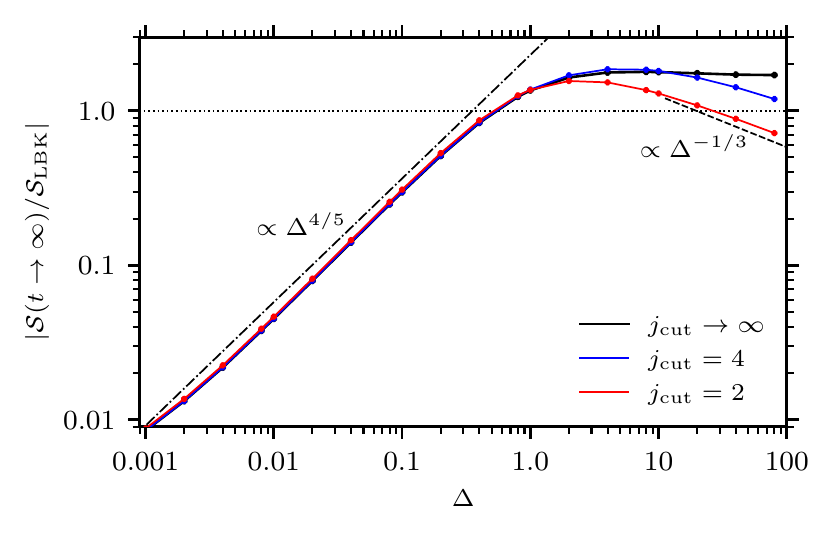}
\caption{Steady-state value of $ \mathcal{S}(t) / \mathcal{S}_\mathrm{LBK}$ as a function of $\Delta$ (note that both axes are on a logarithmic scale).
We show the results for the same choice of resonance location as in Figure \ref{fig:Torque_time_1},
computed with the cutoffs $j_\mathrm{cut} = 2, 4, \infty$.}
\label{fig:Torque_steady}
\end{figure}
To make this analysis more quantitative, in Figure \ref{fig:Torque_steady} we plot
the steady-state value of the slowing rate as a function of $\Delta$, again for $j_\mathrm{cut} = 2, 4, \infty$.
We see that at the low-$\Delta$ end, the steady-state slowing rate 
follows closely the scaling $\propto \Delta^{4/5}$ irrespective of $j_\mathrm{cut}$.
This reflects the fact that for small $\Delta$ the width of the $\mathrm{Im}\,  f_1$ curve is approximately independent of $\Delta$ (Figure \ref{fig:Im_f1}), while its amplitude scales like $\propto \Delta^{4/5}$ (Figure \ref{fig:Max_Values}b).
Meanwhile, at the high-$\Delta$ end in Figure \ref{fig:Torque_steady}, the steady-state slowing rate decays as $\propto \Delta^{-1/3}$ for $j_\mathrm{cut} = 2$ but {approximately} converges towards a constant for $j_\mathrm{cut} \to \infty$.
This can be understood as follows. 
By cutting off the integral in (\ref{eqn:corotation_torque_density}) at $j = \pm 2$ we render the steady-state torque sensitive only to the height, and not the width, of the $\mathrm{Im} \, f_1$ curve, and we know from equations 
(\ref{eqn:Img1})--(\ref{eqn:resonance_function}) and
Figure \ref{fig:Max_Values}b that this height scales like $\Delta^{-1/3}$.  But by extending the domain of integration out to infinity we are able to count all contributions to $\mathrm{Im} \, f_1$; and since we know from \S\ref{sec:asymmetry} that $\int_{-\infty}^\infty \md j\,  \mathcal{R}_\Delta(j) = 1$ is independent of $\Delta$, it is unsurprising that the torque converges for large $\Delta$. {See \S\ref{sec:WeinbergSellwood} for further discussion.}

\begin{figure}
\centering
\includegraphics[width=0.49\textwidth]{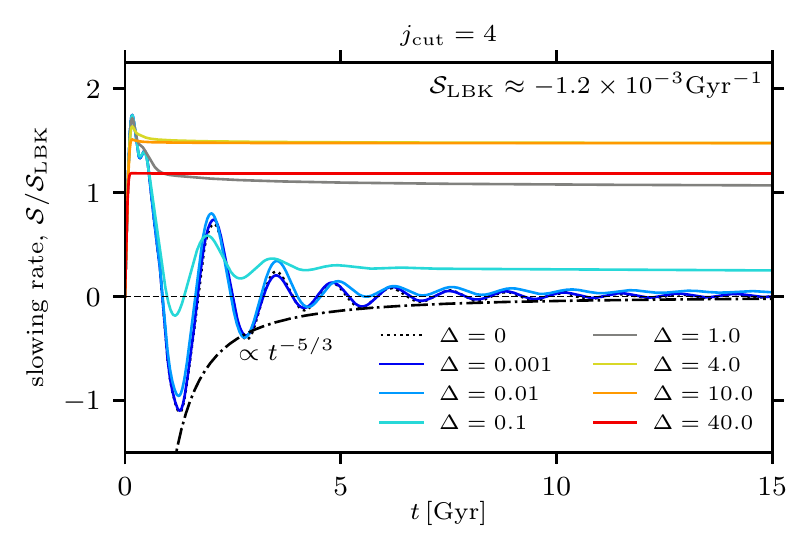}
\caption{As in Figure \ref{fig:Torque_time_1}, except for $J_r = 50$ kpc$^2$ Gyr$^{-1}$ and showing only the 
results for $j_\mathrm{cut} = 4$.}
\label{fig:Torque_time_2}
\end{figure}
In Figure \ref{fig:Torque_time_2} we repeat the same calculation as in Figure \ref{fig:Torque_time_1}b except this time for a much smaller radial action $J_r = 50$ kpc$^2$ Gyr$^{-1}$, 
so the resonance location corresponds to the lower black dot in Figure \ref{fig:Resonant_Line}. The libration timescale in this case is shorter, $t_\mathrm{lib} = 1.3$ Gyr.  Though we are now considering dark matter particles that are typically on much more circular orbits than in Figure \ref{fig:Torque_time_1}b, 
the results we find are remarkably similar. 
The key differences are that
in Figure \ref{fig:Torque_time_2} the amplitude of the 
slowing rate is almost an order of magnitude larger, which reflects the larger value of $\vert L \Psi_1 \vert$ for these more circular orbits, and its oscillation period is shorter owing to the shorter libration time. 
For small $\Delta$, apart from these rescalings there is almost no difference with the results of Figure \ref{fig:Torque_time_1}b.  This again follows from the fact that for weakly-diffusive systems $\mathrm{Im}\,f_1$ tends to be negligible outside of $j\in (-2,2)$.  We can show that the factor $\Psi_1$ does not vary much over this range, meaning it can be approximated by a constant in (\ref{eqn:corotation_torque_density}), and the only remaining $\bJ_\mathrm{res}$ dependence in (\ref{eqn:corotation_torque_density}) lies in the overall prefactor and in the scaling of the time coordinate $\tau = 2\pi t/t_\mathrm{lib}(\bJ_\mathrm{res})$.

Which value of $j_\mathrm{cut}$ is the physically relevant one?
This is actually a nontrivial question, 
since in a detailed calculation one would need to consider all important resonances $\bm{N}$,
 then look carefully at the volume of phase space dominated by the corotation
resonance and choose $j_\mathrm{cut}$ so that the domain of integration does not extend beyond this volume \citep{Chiba2022-qt}.
Moreover, even if corotation was the only important resonance, we could still not fully justify taking $j_\mathrm{cut} \to  \infty$ here because throughout this paper we used the simplification that the background DF $f_0(\bJ)$ could be linearized around the resonance location (equation (\ref{eqn:linear_DF})).
It is clear from Figure \ref{fig:Resonant_Line} that as we venture far from the resonance this 
linear approximation becomes a very poor one.
We could of course generalise our results to more realistic initial conditions \citep{Pao1988-cs}, but this is beyond the scope of this paper.  Also, far from the resonance the pendulum approximation (\ref{eqn:resonant_Hamiltonian}) eventually breaks down.

Nevertheless, regardless of the choice of $j_\mathrm{cut}$ we see from Figure \ref{fig:Torque_steady} that the steady state torque density for $J_r = 500$ kpc$^2$ Gyr$^{-1}$ and $\beta = 0$ 
(the upper resonant location in Figure \ref{fig:Resonant_Line}) is comparable to the LBK value (\ref{eqn:LBK_density}) over the entire range $\Delta \in (1, 100)$, and scales robustly as $\propto \Delta^{4/5}$ for small $\Delta$.
Although not shown here, equivalent plots using $J_r = 50$ kpc$^2$ Gyr$^{-1}$ (the lower resonance in Figure \ref{fig:Resonant_Line}), as well as other choices of $J_r$, show very similar results.
We also note that performing these calculations again for different values of $\beta \neq 0$ gives qualitatively the same results. (However, as mentioned in \S\ref{sec:corotation_torque_density}
the interpretation of such results is more subtle because for $\beta \neq 0$ the orbital inclination precesses,
meaning 
the particles with inclination $\beta$ at time $t$ are not the same particles as were at inclination $\beta$ at time $t - \delta t$.)

\subsubsection{{Summary of this section}}
\label{sec:friction_summary}

{The key result of this section is that the corotation torque density is very sensitive to $\Delta$ at the small-$\Delta$ end, growing almost linearly with $\Delta$, but is much less sensitive
for large $\Delta$, and for $\Delta \gtrsim 1$ the LBK torque is a robust estimate.
More quantitatively, we  suggest the following heuristic fitting formula for the corotation torque density
\begin{equation}
    \mathcal{S}(t\to\infty) \sim \mathrm{min}(\Delta^{4/5}, 1) \times  \mathcal{S}_\mathrm{LBK}.
    \label{eqn:heuristic}
\end{equation}
Alternatively, a reasonable fit to the $j_\mathrm{cut} \to \infty$ curve in Figure \ref{fig:Torque_steady}, accurate to within a few tens of percent over all five decades in $\Delta$, is $\mathcal{S}(t\to\infty) / \mathcal{S}_\mathrm{LBK} = 2\Delta^{4/5}/(1+\Delta^{4/5}$.}

{We discuss the astrophysical implications of our results on bar-halo friction in \S\ref{sec:WeinbergSellwood}.}


\section{Discussion}
\label{sec:Discussion}

Bar resonances are key drivers of the secular evolution of galaxies.
However, analytical studies and test-particle simulations of dynamical interactions 
between a bar and a population of stars or dark matter particles are often highly idealized in the sense that they
are \textit{collisionless}: they only consider the evolution of particle distribution functions (DFs) 
in smooth prescribed potentials, ignoring the diffusive effects of passing stars, molecular clouds, dark matter substructure, transient spiral waves, etc. 
Meanwhile, N-body simulations presumably capture (at least some of) this diffusive physics but 
are difficult to interpret and are rarely linked back quantitatively to the underlying dynamical processes.
Moreover, N-body simulations themselves inevitably include some level of \textit{numerical} diffusion,
even if they purport to
describe a perfectly collisionless system \citep{Weinberg2001-ok,Weinberg2007-bv, Sellwood2013-sr,ludlow2019numerical,Ludlow2021-wk}.

In this paper we have taken a step towards reconciling these various approaches by developing a kinetic framework that accounts for both a secular driving of the DF by a rigidly rotating bar perturbation, and a generic diffusion in the associated slow action variable.
The kinetic equation we proposed is the simplest possible one that allows us to move beyond the paradigmatic collisionless calculations of \citet{Lynden-Bell1972-ve}, \cite{Tremaine1984-wt}, \cite{Binney2016-zh}, etc.
In its dimensionless form (equation (\ref{eqn:kinetic_equation_dimensionless})), the kinetic equation depends on a single dimensionless parameter, the diffusion strength $\Delta$ (equation (\ref{eqn:dimensionless_diffusion})).
All past collisionless models have implicitly set $\Delta = 0$, but in stellar disks we can easily have $\Delta \sim 1$, and $\Delta$ can also be significantly different from zero in the dark matter halo depending on the form that the dark matter takes, suggesting that some of the conclusions drawn from collisionless models may need to be revised.

{In the remainder of this section, we first 
mention various limitations of the model we have developed here (\S\ref{sec:Limitations}).
We then discuss the implications of our results for studies of bar-halo friction (\S\ref{sec:WeinbergSellwood}), 
describe the connection between this work and various seminal studies in plasma physics (\S\ref{sec:Plasma}), and conclude by discussing future directions (\S\ref{sec:Future}).
}

\subsection{{Limitations of this work}}
\label{sec:Limitations}

Our purpose here has not been to provide a detailed model of the Milky Way's (or any other galaxy's) bar-driven evolution, but to elucidate some theoretical ideas upon which future detailed models may be based.
Though we believe we have succeeded in this aim, 
our model is certainly an unrealistic portrayal of any real galactic bar.
For example, 
we have assumed that the bar's resonances are well-separated in phase space so that there is little resonance overlap.  This seems to be a reasonable assumption for dark matter trapping \citep{Chiba2022-qt}, but it is probably not the case for 
stars in the Galactic disk, where resonance overlap is a significant contributor to transport \citep{Minchev2010-la,Minchev2012-ml}.
We have also assumed throughout that the bar is a rigidly rotating structure with constant pattern speed and strength.  Even in our dynamical friction calculations we have not accounted for the fact that the friction causes the pattern speed to change, 
which in turn causes the resonance locations to sweep through action space.  {We discuss this further in \S\ref{sec:WeinbergSellwood}.}
Moreover, many simulations show that the bar parameters actually fluctuate with time \citep{Fujii2018-tt}; recently, \cite{Hilmi2020-kb} simulated two Milky-Way-like galaxies and found that both bar pattern speed and strength fluctuate at the level of $10-20$ percent on orbital timescales ($\sim 100$ Myr), mostly due to interactions with spiral arms. 
{One can show that fluctuations in the
pattern speed and/or strength of the bar causes an effective diffusion of particle orbits (though not precisely as a random walk in $J_\mathrm{s}$).  Heuristically, therefore, one might guess that inclusion of fluctuating bar parameters  amounts to little more than an increased effective $\Delta$. We leave a careful study of this problem to future work}.

Another major assumption we have made throughout the paper is that there is no diffusion of particles' fast actions $\bJ_\mathrm{f}$. 
This assumption is not fully justifiable in the general case --- for example, local isotropically-distributed scattering events should produce diffusion in all three action variables, not just the slow variable $J_\mathrm{s}$.
It is worth mentioning that \textit{the same assumption is made implicitly in collisionless models}.
In fact, in collisionless models, a distinct phase space island
exists at
every $\bJ_\mathrm{f}$ and the DF phase mixes perfectly within this island similar to Figure \ref{fig:slices}a, with no cross-talk between different $\bJ_\mathrm{f}$ values.  This becomes problematic when even a small amount of diffusion is included, because the phase-mixed island structures at each $\bJ_\mathrm{f}$ inevitably produce sharp gradients of $f$ in the $\bJ_\mathrm{f}$ direction, since the location and shape of the resonant island changes as a function of $\bJ_\mathrm{f}$ (for instance $J_\mathrm{s, res}$ itself is a function of $\bJ_\mathrm{f}$ through equation \eqref{eqn:resonance_condition}, and we already argued that the width of the resonance depends on $\bJ_\mathrm{f}$ in \S\ref{sec:corotation_torque_density}).  Because of these sharp gradients, a diffusive term like $\sim \partial^2 f/\partial J_{\mathrm{f}1}^2$ in the kinetic equation is liable to become large, and hence $\bJ_\mathrm{f}$ diffusion will start to `fill in' the flattened regions in the DF produced by phase mixing.\footnote{The longitudinal plasma wave calculation of \cite{Pao1988-cs} \textit{did} include three-dimensional diffusion, but he had the advantage of working in velocity space rather than action space, which simplifies the calculations significantly, and it is not clear whether a comparable angle-action calculation could be developed.}
This suggests that in reality the effective $\Delta$ is again higher than one would naively estimate based on equation
(\ref{eqn:Delta_approximate}).
Nevertheless, our study is still useful in that it allows us to quantify, even if only roughly, the impact of stochastic effects upon
resonant structures through a single intuitive parameter $\Delta$.
If the aforementioned complications do indeed increase the effective $\Delta$, this will only reinforce our main point that in many astrophysically relevant scenarios, diffusive effects
cannot be ignored.

We also neglected to include any drag term in our collision operator (\ref{eqn:collision_operator}), which is usually valid as long as the scattering agents (molecular clouds, spiral arms, etc.) are sufficiently massive compared to the particles in question \citep{Binney1988-zy}.
However, drag can be an important ingredient, for instance if 
$f$ were to describe heavy bodies (e.g. MACHOs) or a population of stars in a tepid disk \citep{Fouvry2015-nk}.
In this case resonance lines, in addition to experiencing broadening, are predicted to shift and split \citep{duarte2020shifting}.

\subsection{Bar-halo friction}
\label{sec:WeinbergSellwood}

{The primary application of our study is to the problem of bar-halo friction (\S\ref{sec:Dynamical_friction}).
This topic has provoked significant controversy in the literature, 
both among theoreticians \citep{Debattista2000-dn,Weinberg2007-bv,sellwood2008bar,athanassoula2013bars} and among simulators and observers, who argue about possible tensions in $\Lambda$CDM cosmology \citep{fragkoudi2021revisiting,roshan2021fast,frankel2022simulated}.
Let us discuss what our results do, and do not, imply about this problem.}

{First, we emphasise that we have not actually calculated the frictional torque at all, but only the 
\textit{corotation torque density} at fixed $J_r$ and $\beta$. Astrophysical inferences from our results are therefore necessarily an extrapolation, although
we take confidence from the fact that our torque density results for $\Delta = 0$ do
resemble closely the full torque calculations of \cite{Chiba2022-qt}.
We have also not considered haloes either with spin or with anisotropic velocity distribution functions, and altering these characteristics may change the bar-halo friction significantly (e.g. \citealt{li2022stellar}).
For instance, the corotation torque will be much reduced in a radially anisotropic halo (compare the LBK
values in Figures \ref{fig:Torque_time_1} and \ref{fig:Torque_time_2}), but the Lindblad resonances might become relatively 
more important.
}

{Despite these major caveats,} suppose we take the formula (\ref{eqn:heuristic}) to be a broadly correct description not only of the corotation torque density, but of the dynamical friction torque as a whole: that is, we replace $\mathcal{S} \to \mathcal{T}$ and reinterpet the (strictly local, $\bJ_\mathrm{f}$-dependent) diffusion strength $\Delta$ as some characteristic value, perhaps averaged
over the important regions of phase space.
 Then our results suggest that real galactic bars will always feel some non-zero torque, 
 since finite diffusion will always replenish some asymmetry in the angular distribution of the particles {as viewed in the bar frame.
Our results also suggest that for systems with moderate to strong diffusion ($\Delta \gtrsim 1$), 
the LBK formula provides a robust order-of-magnitude estimate of the frictional torque.}
Further, for $\Delta \lesssim 1$ the time-asymptotic torque grows almost linearly with $\Delta$, meaning that in some cases a relatively collisional component of a galaxy (e.g. the stellar disk for which $\Delta \sim 1$, see \S\ref{sec:Delta}) might make a significant contribution to the torque, even though it carries much less mass than the nearly-collisionless cold dark matter halo (for which $\Delta \lll 1$).

The fact that the LBK torque formula provides a good order-of-magnitude estimate {for any $\Delta \gtrsim 1$}
can be understood as follows (see also \citealt{Johnston1971-ir}).
Physically, diffusion tends to render the bar-particle interaction problem \textit{linear} in the sense that particles are never truly trapped by the bar --- for instance, they never undergo a full libration around the origin in the $(\phi, j)$ plane (Figure \ref{fig:slices}) before being kicked to a new $j$ value {outside of the resonant island}.
In other words, when diffusion {dominates over libration}, particle trajectories are well-described in the original angle-action variables $(\btheta, \bJ)$, and approximately consist of the unperturbed motion ($\bJ = $ const, $\btheta \propto \bm{\Omega}t$) punctuated by frequent instantaneous jumps in $\bJ$.
This means that one can calculate 
the dynamical friction torque using the same linear perturbation techniques as in the collisionless LBK theory (\S\ref{sec:Dynamical_friction_linear}), while accounting for the additional diffusion.
The effect of this additional ingredient is basically to broaden the resonance line in action space, so that the $\delta$-function encoding exact resonances in the LBK formula (\ref{eqn:LBK_density}) gets replaced by the function $\mathcal{R}_\Delta$ (equation (\ref{eqn:resonance_function})).  But since the `integral under the line' of this broadened resonance function is always unity regardless of $\Delta$, the torque we calculate in the broadened case does not differ much from the LBK result.
More quantitatively, using the identity $\int_{-\infty}^\infty \md x \, p(x) \, \delta\left( q(x) \right) = p(x_0) / \vert q'(x_0) \vert $ where $x_0$ is assumed to be the lone zero of $q(x)$,
we can rewrite the LBK formula (\ref{eqn:LBK_density}) as 
\begin{equation}
    \mathcal{S}_\mathrm{LBK} = (2\pi)^4 \left( L\vert \Psi_1 \vert^2 \frac{\partial f_0}{\partial L} \bigg/  \bigg\vert \frac{\partial \Omega_\psi}{\partial L} \bigg\vert \right)_\mathrm{res},
    \label{eqn:LBK_density_2}
\end{equation}
where the subscript `res' indicates that everything inside the bracket is to be evaluated on resonance.  Next let us crudely approximate the integrand of the nonlinear torque (\ref{eqn:corotation_torque_density}) with its resonant value and `perform the integral' by multiplying this with some resonance width $\delta L$.   Comparing the result with (\ref{eqn:LBK_density_2}), using the fact that for our choice of slow actions
$(\partial \Omega_\psi / \partial L)_\mathrm{res} = G/4$ and $(\partial f_0/\partial L)_{\mathrm{res}} = \alpha / I_\mathrm{h}$, and employing the definitions (\ref{eqn:F_and_G}) and (\ref{eqn:half_width}), one can show that
\begin{equation}
  \frac{\mathcal{S}(t\to \infty)}{ \mathcal{S}_\mathrm{LBK}} \approx \frac{2}{\pi} \frac{\delta L}{I_\mathrm{h}} \mathrm{Im} \, g_1(0).
    \end{equation}
We know from \S\ref{sec:asymmetry} that for large $\Delta$, the resonance width is $\delta L \sim \Delta^{1/3} I_\mathrm{h}$ and $\mathrm{Im}\,g_1(0) \sim \Delta^{-1/3}$. 
It follows that as long as $\Delta \gtrsim 1$ we always have $
\mathcal{S}(t\to \infty)/ \mathcal{S}_\mathrm{LBK} \sim 1$.

{Note that the above argument does \textit{not} hold in the $\Delta < 1$ regime, in which trapped orbits typically do undergo at least one full libration before being kicked out of resonance.  In this regime the  `resonance function' takes on a qualitatively different form --- indeed, one can see from Figure \ref{fig:Im_f1} that the `area under the curve' of $\mathrm{Im}\, g_1$ is not close to a constant for small $\Delta$ values.
}

{
Similar conclusions to these
were also reached by \cite{Weinberg2007-bv}, who 
considered the effect of finite particle number upon simulations of bar-halo coupling.
They derived criteria for the minimum particle number $N_\mathrm{crit}$ required such that a simulation
might reproduce faithfully the collisionless (TW84) dynamics, rather than destroying the nonlinear trapping effects by spurious numerical diffusion\footnote{As well as satisfying certain `numerical coverage' criteria which we will not
discuss here.}.  In our language, they were trying to determine the minimum number of particles $N$ that would still produce $\Delta \ll 1$ (see equation \eqref{eqn:Delta_N}).
}
{However, as \cite{Weinberg2007-bv} noted, the time-asymptotic TW84 limit \eqref{eqn:TW84} may never be reached even in a genuinely collisionless $(\Delta = 0)$ system if the pattern speed of the bar, $\pattern$, is allowed to change self-consistently, since this will cause the resonance location to sweep through phase space.
In particular, if $\pattern$ changes fast enough that the resonance sweeps past new halo
material on a timescale short compared to $t_\mathrm{lib}$ --- the so-called `fast limit', see TW84 --- then nonlinear librations never occur,
and the torque on the bar is well approximated by linear (LBK) theory\footnote{Or rather, the time-dependent version of LBK theory --- see equation \eqref{eqn:Torque_linear} and \cite{Weinberg2004-ss}.}.}
{On the other hand, this trend can be halted if $\pattern$ evolves non-monotonically \citep{sellwood2006bar}.  In particular, if a gently-slowing bar experiences a small positive fluctuation in $\pattern$, the resonances responsible for friction will be nudged `back' to the phase space locations they just came from.  At these locations, the gradient in the distribution function will have been flattened.  This will lead to an anomalously small frictional torque relative to the LBK value, helping to keep the bar  the slow regime (what \citealt{sellwood2006bar} call a `metastable' state).}

{Two key points emphasized by \cite{Weinberg2007-bv} were as follows. (i) Even if the real system one wishes to simulate genuinely does lie in the `slow limit' where nonlinear trapping is crucial, 
a finite-$N$ simulation of this system with too much numerical diffusion may produce evolution in the fast limit.
Our results are consistent with this idea: increasing the diffusion strength $\Delta$ from a small value $\ll 1$ tends to increase the frictional torque.
 (ii) If a simulation employs far too few particles, $N \ll N_\mathrm{crit}$, then
a simple convergence study in $N$ may be misleading.  In the language of our paper, the reason for this
is that if the initial simulation had $\Delta \gg 1$, then e.g. 
a ten-fold increase in $N$ will only lead to a ten-fold decrease in $\Delta$. If such a decrease still leaves $\Delta > 1$
then we know from Figure \ref{fig:Torque_steady} that one will measure 
 almost precisely the same torque again, so the numerics will appear converged.
In other words, one may need to increase $N$ much further,
with no apparent change in the measured torque
$\mathcal{T}(N)$
until it eventually
changes discontinuously around $N\sim N_\mathrm{crit}$ (corresponding to $\Delta \sim 1$).  
Our Figure \ref{fig:Torque_steady}
helps to illustrate and quantify both points (i) and (ii), but a proper study with
time-dependent $\pattern$ and a very wide range of effective particle numbers
is required to make these claims precise \citep{sellwood2008bar}.
} {In particular, the mechanism from \cite{sellwood2006bar} described above is a particularly sensitive test case; the ability of a simulation to reproduce this effect would be a reassuring sign (though not a guarantee) that particle number requirements are being met.}

{Finally, we should not forget that there are many decades in $\Delta$ 
that lie between the TW84 ($\Delta = 0$) and LBK ($\Delta \gtrsim 1$) limits, 
especially since it is in this regime that most cosmological simulations probably sit (\S\ref{sec:Delta}).  Indeed, even $\Delta \sim 0.1$ can lead to a significant bar slowdown over a Hubble time, and such values are plausible even in simulated haloes that are naively converged, i.e. those with two-body relaxation times $\gg 10$ Gyr.
Further work along the lines of \cite{Sellwood2013-sr,ludlow2019numerical,Ludlow2021-wk,wilkinson2023impact}
will
be needed to confirm the extent to which bar slowdown is driven by spurious numerical noise.
Our results also suggest that alternative dark matter models such as fuzzy dark matter \citep{Hui2017-rw}
may produce different 
bar-halo friction signatures, potentially constraining 
cosmological models (e.g. \citealt{Debattista1998-nv,Lancaster2020-vt}).}

\subsection{Relation to plasma literature}
\label{sec:Plasma}

\begin{deluxetable*}{ccccc}
\label{table:literature}
\tablenum{1}
\tablecaption{Guide to some complementary papers in the analytic theory of wave-particle interactions in plasma kinetics (shown in regular text) and galactic dynamics (shown in bold).  We split the papers by whether the authors considered the linear or the nonlinear response of the particle DF to the wave perturbation, and whether their calculation did or did not include diffusion. The present paper belongs in the lower-right quadrant.}
\tablewidth{0pt}
\tablehead{
\nocolhead{\textcolor{red}{plasma} / \textbf{galactic}} & \colhead{\textit{Collisionless}} & & \multicolumn2c{\textit{Collisional}}}
\startdata
 & {\citet{Landau1946-aj}} & & \multicolumn2c{\citet{Auerbach1977-xe}}  \\
\textit{Linear theory} & \textbf{\citet{Lynden-Bell1972-ve}} & & \multicolumn2c{\citet{Catto2020-qo}}\\
 & \textbf{\citet{Weinberg2004-ss}} & &\\
 \hline
 & \colhead{\fbox{$\Delta = 0$}} & & \colhead{\fbox{$0 < \Delta \ll 1$}} & \colhead{\fbox{$\Delta \gg 1$}}
 \\
 \textit{Nonlinear theory} & \cite{ONeil1965-uy,mazitov1965damping} &  & {\citet{Pao1988-cs}} & {\citet{BerkPPR1997}}   \\
\textit{(w/particle trapping)} & \textbf{\citet{Tremaine1984-wt}} & & {\citet{Petviachvili1999-mc}} & {\citet{Duarte2019-pl}}  \\
 & \textbf{\citet{Chiba2022-qt}} & & \multicolumn2c{\textbf{$<$---  this paper (Hamilton et al. 2023)  ---$>$}} \\
\enddata
\end{deluxetable*}

A note is warranted here on the many correspondences between stellar-dynamical and plasma-kinetic theory.  
Our kinetic equation (\ref{eqn:kinetic_equation_dimensionless}) turns out to be mathematically identical to an equation employed in a range of 
plasma kinetic calculations of wave-particle interactions \citep{Pao1988-cs,BerkPPR1997,Duarte2019-mi}.
However we have not seen it expressed in this dimensionless single-parameter form before, nor have we seen the solutions examined so closely as in \S\ref{sec:solution},
meaning our results may be of use for future plasma studies.
The plasma literature more generally is replete with analyses of nonlinear particle trapping, diffusive resonance broadening, and so on (e.g. \citealt{Dupree1966-dl,Su1968-ga,Ng2006-gu,Black2008-cg,White2019-nl,Catto2020-qo,Catto2021-ed,Tolman2021-ly}).
Moreover, 
the LBK formula \eqref{eqn:LBK} is directly analogous to the classic formula for the Landau damping rate of a Langmuir wave in an electrostatic plasma \citep{Landau1946-aj,Ichimaru1965-zm} or Landau damping in more general geometry \citep{Kaufman1972-ag,Nelson1999-in}.  The TW84 calculations that account for
nonlinear trapping of particles in resonances are strongly analogous to the `nonlinear Landau damping' calculations by \cite{ONeil1965-uy,mazitov1965damping} (who, unsurprisingly, found that the damping/growth rate of waves goes to zero in the fully phase-mixed limit).
In a similar vein, the calculations in the present paper are analogous to those works that have attempted to combine O'Neil's and Mazitov's calculation with
a simple model for inter-particle collisions (e.g. \citealt{Zakharov1963-ae,Pao1988-cs,Brodin1997-sj}).
The approximate recovery of the LBK torque from the nonlinear torque in the large $\Delta \gg 1$ limit is closely related to the recovery of the Landau damping formula from the O'Neil/Mazitov formula \citep{Johnston1971-ir,Auerbach1977-xe}.  In Table \ref{table:literature} we provide a summary guide to this plasma-kinetic/stellar-dynamical correspondence, and show where the present paper fits into the literature.

\subsection{Outlook and future work}
\label{sec:Future}

In recent years our analytic understanding of dynamical friction in the collisionless ($\Delta=0$) limit has been greatly improved e.g. by the work of \cite{Chiba2022-qt}, \cite{Banik2021-ug,Banik2022-rj} and \citet{Kaur2018-wp,Kaur2022-tb}.
To some extent these authors have advocated an `orbit-based' interpretation of dynamical friction.  For instance, \cite{Chiba2022-qt} and \cite{Banik2022-rj} calculated the contributions to the friction from individual particles at different points in their orbits, compared the torque coming from trapped orbits to those from untrapped orbits, and so on.
Our finite-$\Delta$ calculations show the benefit of a complementary viewpoint, in which dynamical friction is a kinetic process --- something that needs to be understood primarily in terms of distribution functions as opposed to individual particle trajectories.
In reality, no particle is permanently trapped or untrapped by a resonance, since for finite $\Delta$ there is a constant flux of particles into and out of the librating region even in steady state. 

{The most obvious extension of the current work is to a system in which the bar pattern speed
$\pattern$ is decaying monotonically \citep{Weinberg2007-bv,Weinberg2007-zo,sellwood2008bar}. 
We must also quantify carefully the impact of diffusion in fast actions 
$\bJ_\mathrm{f}$, and calculate the full torque by integrating over $\bJ_\mathrm{f}$, to 
see whether a single scalar ($\Delta$) parameterization is useful 
in the fully two- or three-dimensional case.
More broadly, by understanding the time-dependent problem including diffusion,
we should be able to reconcile the various theoretical, numerical and
observational approaches to 
bar-halo friction, which is one of the most promising small-scale windows onto the nature of dark matter.
An investigation along these lines is underway.
}


\section{Summary}
\label{sec:Summary}

In this paper we have investigated the impact of diffusion upon the resonant imprints left by rigidly rotating galactic bars in the distribution function (DF) of stars and dark matter particles, and the subsequent effect this has on {bar-halo friction}.  Our key findings can be summarised as follows.

\begin{itemize}
    \item Using the pendulum approximation to describe the secular forcing by the bar and assuming a simple diffusion
    in the associated slow action variable, we proposed a
     kinetic equation for the DF that depends on just one parameter, the diffusion strength $\Delta$, equal to 
    the ratio of the resonant libration period to the diffusion time across the resonance. Though many classic analytic studies took $\Delta = 0$ by default, $\Delta$ can in fact be significantly larger than zero {both in real astrophysical systems, and in simulations of supposedly collisionless systems, \textit{even if the system's naive relaxation time is a Hubble time or longer}.} 
    \item Assuming simple initial/boundary conditions we solved this kinetic equation for a wide range of $\Delta$ values.
    For $\Delta =0$ and $t\to \infty$ we recovered the classic result in which the DF is spread uniformly (`phase-mixed') along the pendulum Hamiltonian contours in slow angle-action space.
    This phase-mixed structure was broken for finite $\Delta$, and we provided an analytic understanding of the resulting DF in the $\Delta \ll 1$ and $\Delta \gg 1$ limits.
    \item We calculated the \textit{corotation torque density} $\mathcal{S}$ felt by a galactic bar due to its frictional interaction with an initially spherical halo of dark matter particles, focusing on the contribution from particles orbiting coplanar with the bar, for various $\Delta$.
    For $\Delta =0$ we recovered the result of \cite{Tremaine1984-wt} that $\mathcal{S} \to 0$ as $t\to \infty$. However,
    since diffusion replenishes the asymmetry of the DF in the resonant region of phase space,
    the steady-state torque never vanished for finite $\Delta$. 
    Instead we found $\mathcal{S} \sim \mathrm{min}(\Delta^{4/5}, 1) \times \mathcal{S}_\mathrm{LBK}$, where $\mathcal{S}_\mathrm{LBK}$  is the (negative) linear theory result from \cite{Lynden-Bell1972-ve}.
\end{itemize}
Our work highlights the fact that resonant phenomena are \textit{delicate}, and that one must therefore be mindful of the effects of diffusion if one is to understand how resonances sculpt galaxies.  We have shown this for the particular scenario of bar-dark matter halo coupling, 
but we suspect that {analogous} 
conclusions {may} apply to resonant imprints left in Solar neighborhood kinematic data. 
Further work will be needed to make our rather qualitative claims precise.

\begin{acknowledgements} 

We are grateful to the referee for a close reading and several suggestions that strengthened our paper considerably.
We also thank S.~Tremaine, C.~Terquem, R.~Rafikov, R.~Sanderson, M.~Weinberg, K.~Johnston, B.~Kocsis, T.~Yavetz, U.~Banik, E.~Vasiliev, S.~Chakrabarti, C.~Pichon, A.~Bhattacharjee, R.~Chiba, L.~Beraldo e Silva, and members of the CCA Dynamics group for helpful comments and discussions.
This work was supported by a grant from the Simons Foundation (816048, CH), by the U.S. Department of Energy under contract DE-AC02-09CH11466 (VND), by the Institute for Advanced Study (LA), and by the W.M.~Keck Foundation Fund at
the Institute for Advanced Study (ET). This work was performed in part at Aspen Center for Physics, which is supported by National Science Foundation grant PHY-1607611.

\end{acknowledgements}


\appendix

\section{Analytic results for weak and strong diffusion}

In this Appendix we derive some analytic results that allow us to understand the
solutions to the kinetic equation (\ref{eqn:kinetic_equation_dimensionless}) 
in the asymptotic limits of weak diffusion $\Delta \ll 1$ (\S\ref{sec:weak_scattering}) and the strong diffusion $\Delta \gg 1$ (\S\ref{sec:strong_scattering}).


\subsection{The limit of weak diffusion, $\Delta \ll 1$}
\label{sec:weak_scattering}

\subsubsection{Phase-mixed DF in the collisionless limit $\Delta =0$}

The (coarse-grained) steady-state phase-mixed DF for $\Delta =0$ is easily calculated by smearing the initial DF evenly over the phase space contours, as described for example by \cite{Binney2016-zh}, \cite{Monari2017-tu} and, in a different context, by \cite{Hamilton2022-gr}.  
Let us call this phase-mixed DF $f_\mathrm{pm}(\phi, j)$; then we know it will only be a function of $h$, where
\begin{equation}
    h(\phi, j) \equiv \frac{1}{2}j^2 - \cos \phi. 
    \label{eqn:dimensionless_H}
    \end{equation}
    is the effective dimensionless Hamiltonian. 
    The initial DF $f_\mathrm{init}$ (equation \eqref{eqn:linear_DF}) written in terms of  $(\phi, h)$ is
\begin{equation}
    f_\mathrm{init} = f_0(0) + \alpha  j_\pm(\phi, h),
    \label{eqn:f_init_h}
\end{equation}
where $j_\pm \equiv \pm \sqrt{2(h+\cos \phi)}$ and we take $+$ or $-$ depending on whether 
we are applying the equation to positive or negative $j$.
Now, for $\Delta = 0$ there are two well-defined orbit families, namely librating ($h < 1$) and circulating ($h>1$) orbits, distinguished by a separatrix ($h=1$).
Since the resonant island is symmetric around $j=0$ (e.g. Figure \ref{fig:dimensionless_DF}), it is geometrically obvious that for librating orbits the average phase space density on any given contour is just
\begin{equation}
    f^\mathrm{lib}_{\mathrm{pm}} = f_0(0),
    \label{eqn:phase_mixed_librating}
\end{equation}
and hence $g^\mathrm{lib}_{\mathrm{pm}}=0$ (see the final panel of Figure \ref{fig:slices}a).
For circulating orbits ($h>1$) we use the fact that 
the line element in the $(\phi_*, j_*)$ plane at fixed $h$ is
\begin{eqnarray}
    \md \lambda = 
    \md \phi_* \sqrt{1+ \frac{\sin^2\phi_*}{j^2_\pm(\phi_*, h)}}.
\end{eqnarray}
The phase-mixed DF at fixed $h(\phi, j)$ is then found by averaging \eqref{eqn:f_init_h} over the contour: 
\begin{eqnarray}
    f_{\mathrm{pm}}^\mathrm{circ}(h(\phi,j)) = f_0(0) + \alpha \frac{\oint  \md \lambda \,  j_\pm(\phi_*, h)}{\oint  \md \lambda}.
    \label{eqn:phase_mixed_ratio}
\end{eqnarray}
Equations (\ref{eqn:phase_mixed_librating})--(\ref{eqn:phase_mixed_ratio}) describe fully the phase-mixed DF $f_\mathrm{pm}$, towards which the solution to (\ref{eqn:kinetic_equation_dimensionless}) tends in the time-asymptotic limit.

\subsubsection{Steady-state DF for finite diffusion, $0 < \Delta \ll 1$}

When diffusion is finite but weak ($0 < \Delta \ll 1$), the steady-state solution to (\ref{eqn:kinetic_equation_dimensionless}) is modified somewhat from the perfectly phase-mixed solution $f_\mathrm{pm}$ --- see Figure \ref{fig:slices}b and the blue lines in Figures \ref{fig:phi_averaged_g}c and \ref{fig:Im_f1}.
We can gain some insight into this modified steady-state DF 
following the method of \cite{Pao1988-cs} (see also \citealt{Petviachvili1999-mc}).
%
%
%
%

First, we look for a steady-state perturbative solution
\begin{equation}
    f_\mathrm{steady}(\phi, j) = f_\mathrm{pm}(h) + \delta f_\mathrm{steady}(\phi, j),
    \label{eqn:small_Delta_ansatz}
\end{equation}
where $\vert \delta f_\mathrm{steady}/f_\mathrm{pm}\vert  = \mathcal{O}(\Delta)$. 
We then have from \eqref{eqn:kinetic_equation_dimensionless} that
\begin{equation}
   j \frac{\partial \delta f_\mathrm{steady}}{\partial \phi} - \sin \phi \frac{\partial \delta f_\mathrm{steady}}{\partial j} \approx \Delta \frac{\partial^2 f_\mathrm{pm}}{\partial j^2}.
   \label{eqn:steady_state_small_Delta_perturbation}
\end{equation}
To solve this equation we change variables from $(\phi, j) \to (\phi, h)$, where $h$ is given in (\ref{eqn:dimensionless_H}).  Then equation \eqref{eqn:steady_state_small_Delta_perturbation} becomes
\begin{equation}
   \frac{\partial \delta f_\mathrm{steady}}{\partial \phi} 
   = \Delta \frac{\partial }{\partial h} \left[ j_\pm(\phi, h) f_\mathrm{pm}'(h)\right].
   \label{eqn:perturbation_new_coords}
\end{equation}
The solution to \eqref{eqn:perturbation_new_coords} is
\begin{eqnarray}
    \delta f_\mathrm{steady}(\phi, h)  = \pm \Delta \frac{\partial}{\partial h} \left[ f_\mathrm{pm}'(h) \int^\phi \md x \sqrt{2(h+\cos x)}\right]
    = \pm  \Delta \frac{\partial}{\partial h} \Bigg[
    2\sqrt{2(1+h)} f_\mathrm{pm}'(h) E\left( \frac{\phi}{2}, \sqrt{\frac{2}{1+h}} \right)\Bigg],
    \label{eqn:delta_f_weak_scattering_full}
\end{eqnarray}
where $E(\phi,x)\equiv \int_0^\phi \md y \sqrt{1-x^2 \sin^2 y}$ is an incomplete elliptic integral of the second kind, and 
we take $+$ or $-$ depending on whether we apply the equation to $j>0$ or $j<0$.
For the circulating region of phase space ($h>1$) we can get an explicit form for $f_\mathrm{pm}'(h)$ as follows. 
We integrate equation \eqref{eqn:perturbation_new_coords} with respect to $\phi$ from $-\pi$ to $\pi$ and get
\begin{eqnarray}
0 = \frac{\partial}{\partial h} \left[ 4 {f^\mathrm{circ}_\mathrm{pm}}'(h) \sqrt{h-1} 
E\left(\sqrt{\frac{-2}{h-1}}\right)\right],
\label{eqn:delta_f_small_Delta}
\end{eqnarray}
where $E(x) \equiv E(\pi/2, x)$ is the complete elliptic integral of the second kind.
It follows that the quantity in square brackets is equal to a constant.
We can fix this constant by using the fact that very far from the resonance we have $h \approx j^2/2 \gg 1$ and $f_\mathrm{pm}$ is unperturbed, $f_\mathrm{pm} \approx f_\mathrm{init} = f_0(0) \pm \alpha \sqrt{2h}$ (see equation (\ref{eqn:linear_DF})).
Then we find that the square bracket is equal to $\pm \pi \sqrt{2} \alpha$, so\footnote{Of course one can also derive an exact expression for ${f^\mathrm{circ}_\mathrm{pm}}'(h)$ by differentiating the phase mixed solution \eqref{eqn:phase_mixed_ratio}, but the approximate expression \eqref{eqn:phase_mixed_derivative} is simpler and sufficiently accurate for our purposes.}
\begin{equation}
    {f^\mathrm{circ}_\mathrm{pm}}'(h) = \frac{\sqrt{2}\pi\alpha}{4\sqrt{h-1}} \left[ E\left(\sqrt{\frac{-2}{h-1}}\right) \right]^{-1}.
    \label{eqn:phase_mixed_derivative}
\end{equation}
Combining this with \eqref{eqn:delta_f_weak_scattering_full} we have
that for $h>1$,
\begin{eqnarray}
    \delta f^\mathrm{circ}_\mathrm{steady}(\phi, h)  = \pm  \alpha \Delta \pi  \frac{\partial}{\partial h} \Bigg[
    \sqrt{\frac{h+1}{h-1}}\frac{E\left( \phi/2,\sqrt{ 2/[1+h] }\right) }{E\left(\sqrt{-2/[h-1]}\right)}
    \Bigg] .
    \label{eqn:delta_f_weak_scattering_circulating}
\end{eqnarray}
Though we do not illustrate it here, equation (\ref{eqn:delta_f_weak_scattering_circulating}) does a fairly good job of reproducing the numerical solution for the region $\vert j\vert > 2$.
On the other hand, when we apply the same solution technique to the 
librating portion of phase space ($h<1$), it fails.
The reason is that $f^\mathrm{lib}_\mathrm{pm}(h)=$ constant (see equation \eqref{eqn:phase_mixed_librating}) so that ${f^\mathrm{lib}_\mathrm{pm}}'(h) = 0$, and hence \eqref{eqn:delta_f_weak_scattering_full} would predict $\delta f^\mathrm{lib}_\mathrm{steady} = 0$.  This a poor solution, as can be seen by inspecting Figure \ref{fig:slices}b.
Of particular importance for dynamical friction calculations (\S\ref{sec:Dynamical_friction}) is the fact that
this solution fails to reproduce the major contributions to the $\mathrm{Im}\,  g_1$ curve for $\vert j \vert < 2$ shown for small $\Delta$ in Figure \ref{fig:Im_f1}.

The reason for the failure of this solution is that for small $\Delta$, sharp gradients in the DF tend to build up near the separatrix, meaning that the term $\Delta \partial^2 \delta f_\mathrm{steady}/\partial j^2$ on the right hand side of equation \eqref{eqn:kinetic_equation_dimensionless}
cannot be neglected --- the approximation \eqref{eqn:steady_state_small_Delta_perturbation} therefore breaks down. 
To remedy this one can treat the region near the separatrix as a boundary layer where $\delta f_\mathrm{steady}$ can be of the same order as $f_\mathrm{pm}$.
The reader is referred to \cite{Pao1988-cs} for details of how this calculation works, 
although in practice it is a cumbersome task and the final expressions are rather unenlightening. 
The important point is that one gets an additional skew-symmetric contribution to the DF (in the sense of (\ref{eqn:skew})) in a layer on the inner edge of the separatrix, which is of the same order as the phase-mixed solution $f_\mathrm{pm}$. The thickness of the layer grows
with $\Delta$ although not in a way that is easily quantified; empirically this leads to the low-$\Delta$ behavior described in \S\ref{sec:solution}, and in particular to the approximate scaling $\mathrm{max}(\mathrm{Im}\,g_1) \propto \Delta^{4/5}$ (Figure \ref{fig:Max_Values}b).

\subsection{The limit of strong diffusion, $\Delta \gg 1$}
\label{sec:strong_scattering}

Let us now consider the opposite limit, in which the timescale for diffusion $t_\mathrm{diff}$ is short compared to the libration time $t_\mathrm{lib}$, i.e. $\Delta \gg 1$. We note that the 
initial DF (\ref{eqn:linear_DF}) is annihilated by the collision operator on the right hand side of (\ref{eqn:kinetic_equation_dimensionless})
because  $\Delta\, \partial^2 f_\mathrm{init}(j)/\partial j^2 = 0$. 
Thus in the limit of infinitely strong diffusion $\Delta \to \infty$, when this term is the dominant one in the kinetic equation, we expect that the solution $f(\phi, j,  \tau)$ will never deviate from $f_\mathrm{init}$. Therefore, without loss of generality let us write
\begin{eqnarray}
    f(\phi, j, \tau) = f_\mathrm{init}(j) + \delta f(\phi, j, \tau).
    \label{eqn:strong_ansatz}
\end{eqnarray}
Plugging this into the kinetic equation (\ref{eqn:kinetic_equation_dimensionless}) we find
\begin{equation}
    \frac{\partial \delta f}{\partial \tau} + j\frac{\partial \delta f}{\partial \phi} - \sin \phi \left( \alpha + \frac{\partial \delta f}{\partial j}\right)= \Delta \frac{\partial^2 \delta f}{\partial j^2}.
    \label{eqn:kinetic_equation_linearized_full}
\end{equation}
For large but finite diffusion, $\Delta \gg 1$, we anticipate that $\delta f$ will be finite but small (see Figure \ref{fig:slices}d), so that $\epsilon(\phi, j) \equiv \vert (\partial \delta f/\partial j) /\alpha \vert \ll 1$ is small.  Our aim in this section will be to compute $\delta f$ to lowest order in this small parameter.  Higher order solutions are also possible \citep{Duarte2019-mi} but will not be necessary for our purposes. Let us therefore drop the term $\partial \delta f/\partial j$ from \eqref{eqn:kinetic_equation_linearized_full} and write
\begin{equation}
    \frac{\partial \delta f}{\partial \tau} + j\frac{\partial \delta f}{\partial \phi} -  \alpha \sin \phi \approx \Delta \frac{\partial^2 \delta f}{\partial j^2}.
    \label{eqn:kinetic_equation_linearized}
\end{equation}
We now take advantage of the periodicity of the coordinate system to expand $\delta f$ as a Fourier series: $
    \delta f(\phi, j, \tau) = \sum_{n = -\infty}^\infty \delta f_n(j,\tau) \exp(in\phi)$. Since $f$ is real we must have $\delta f_n^{*}(j,\tau)=\delta f_{-n}(j,\tau)$. Plugging the Fourier expansion in to (\ref{eqn:kinetic_equation_linearized})
and using the orthogonality of Fourier harmonics, we get 
\begin{equation}
    \frac{\partial \delta f_n}{\partial \tau} + in j \, \delta f_n
    + \frac{i \alpha}{2} (\delta_{n-1}^0 - \delta_{n+1}^0)
     = \Delta \frac{\partial^2 \delta f_n}{\partial j^2}.
    \label{eqn:kinetic_equation_Fourier}
\end{equation}
The solution to this equation is $\delta f_n = 0$ for $n\neq \pm 1$, {while for $n=\pm 1$ we have}:
\begin{equation}
    \delta f_{\pm 1}(j, \tau) =  \pm \frac{\alpha}{2i{\Delta^{1/3}}}\int_0^{{\Delta^{1/3}}\tau} \md y \, \exp\left(-\frac{y^3}{3} \mp \frac{ijy}{{\Delta^{1/3}}} \right).
    \label{eqn:fourier_solution}
\end{equation}
It follows that the solution to the kinetic equation (\ref{eqn:kinetic_equation_dimensionless}) for $\Delta \gg 1$ is approximately
\begin{eqnarray}
    f(\phi, j, \tau) = f_\mathrm{init}(j) + 
    \frac{ \alpha}{{\Delta^{1/3}} }
 \int_0^{{\Delta^{1/3}} \tau} \md y \sin\left( \phi - \frac{jy}{{\Delta^{1/3}}}\right)\exp\left(-\frac{y^3}{3}\right).
\label{eqn:delta_f_1st_order_solution}
\end{eqnarray}
The $\exp(-y^3/3)$ factor in the integrand of (\ref{eqn:delta_f_1st_order_solution})
means that the contribution to the integral from $y$ values $\gtrsim 2$ are negligible, meaning the steady state is approximately reached when $\tau \sim \Delta^{-1/3}$.

We note also that in this limit we get an explicit expression for the key quantity $\mathrm{Im} f_1$ in steady state:
\begin{equation}
       \mathrm{Im} \, f_{1}(j, \tau \to \infty) =  \frac{\alpha}{2{\Delta^{1/3}}}\int_0^{\infty} \md y \, \exp\left(-\frac{y^3}{3}\right) \cos\left(\frac{jy}{{\Delta^{1/3}}} \right),
\end{equation}
which leads directly to equations (\ref{eqn:Img1})--(\ref{eqn:resonance_function}).

Finally, we can use the solution (\ref{eqn:delta_f_1st_order_solution}) to check our assumption that $\epsilon \equiv \vert (\partial \delta f/\partial j) /\alpha \vert \ll 1$.  We take the derivative of $f-f_\mathrm{init}$  with respect to $j$ and divide by $\alpha$ to find
\begin{eqnarray}
    \epsilon &\approx& \Delta^{-2/3}\bigg\vert \int_0^{\infty} \md y \, y\sin\left( \phi - \frac{jy}{{\Delta^{1/3}}}\right)\exp\left(-\frac{y^3}{3}\right) \bigg\vert
    \nn
    \\
    &\leq& \Delta^{-2/3}\bigg\vert \int_0^{\infty} \md y \, y \exp\left(-\frac{y^3}{3}\right) \bigg\vert \approx  0.94\Delta^{-2/3}.
\end{eqnarray}
Thus $\epsilon$ is small as long as $\Delta \gtrsim$ a few.

\vspace{5mm}






\bibliography{sample631}{}
\bibliographystyle{aasjournal}



\end{document}